\documentclass[12pt]{article}
\usepackage{amsmath}
\usepackage{geometry}
\usepackage{graphicx}
\usepackage{amssymb}
\usepackage{epsfig}

\newif{\ifcomentarios}
\comentariosfalse

\newtheorem{theorem}{Theorem}

\newtheorem{corollary}[theorem]{Corollary}

\newtheorem{definition}[theorem]{Definition}

\newtheorem{lemma}[theorem]{Lemma}

\newtheorem{proposition}[theorem]{Proposition}
\newtheorem{remark}[theorem]{Remark}

\begin{document}

\author{Alexei M. Veneziani\thanks{%
The asymptotic analysis employed in the present work was presented as part
of this author's PhD thesis at the University of S\~{a}o Paulo, supported by
FAPESP under grant \# 02/10954-7. }\ \thanks{
E-mail: \texttt{alexei.veneziani@ufabc.edu.br} }\ \ \ ,\ \ \ \ Tiago Pereira%
\thanks{%
E-mail: \texttt{tiago.pereira@ufabc.edu.br} } \\
Centro de Matem\'{a}tica, Computa\c{c}\~{a}o e Cogni\c{c}\~{a}o \\
Universidade Federal do ABC\\
Av. dos Estados 5001, Santo Andr\'{e}, SP, Brasil \ \\
and \and Domingos H. U. Marchetti\thanks{%
Present Address: Mathematics Department, \ The University of British
Columbia, Vancouver, \ BC, \ Canada \ V6T 1Z2. Email: \texttt{%
marchett@math.ubc.ca}} \\
Instituto de F\'{\i}sica\\
Universidade de S\~{a}o Paulo \\
Caixa Postal 66318\\
05315 S\~{a}o Paulo, SP, Brasil}
\title{Asymptotic Integral Kernel for Ensembles of Random Normal Matrix with
Radial Potentials}
\date{}
\maketitle

\begin{abstract}
We use the steepest descents method to study the integral kernel of a family
of normal random matrix ensembles with eigenvalue distribution%
\begin{equation*}
P_{N}(z_{1},\cdots ,z_{N})=Z_{N}^{-1}\emph{e}^{-N\sum_{i=1}^{N}V_{\alpha
}(z_{i})}\prod_{1\leq i<j\leq N}\left\vert z_{i}-z_{j}\right\vert ^{2}
\end{equation*}%
where $V_{\alpha }\left( z\right) =\left\vert z\right\vert ^{\alpha },$ $%
z\in \mathbb{C}$ and $\alpha \in \left] 0,\infty \right[ $. Asymptotic
analysis with error estimates are obtained. A corollary of this expansion is
a scaling limit for the $n$-point function in terms of the integral kernel
for the classical Segal--Bargmann space.

\medskip


\medskip

\end{abstract}

\newpage

\section{Introduction and statement of the main result}

\setcounter{equation}{0} \setcounter{theorem}{0}

The investigation of non--Hermitian random matrices, whose elements are
independent complex Gaussian variables without any constraint, began with
the work of Ginibre \cite{Ginibre}. Applying the theory of Haar measure to
the group $GL\left( N,\mathbb{C}\right) $ of $N\times N$ complex matrices,
the joint probability distribution of the eigenvalues has shown to be given
by (\ref{P1}) with $V(z)=\left\vert z\right\vert ^{2}$ and the eigenvalue
density in the complex plane, defined by 
\begin{equation*}
\int_{A}\rho _{N}(z)d^{2}z=\frac{1}{N}\mathbb{E}\left( \#\left\{ \text{%
eigenvalues in }A\right\} \right)
\end{equation*}%
for any Borel set $A\subset \mathbb{C}$, where $\mathbb{E}\left( \cdot
\right) $ is the expectation with respect to $P_{N}$, has shown to converges
to the so called circular law%
\begin{equation}
\rho (z)=\left\{ 
\begin{array}{ll}
\dfrac{1}{\pi } & \mathrm{if\ }\left\vert z\right\vert \leq 1 \\ 
0 & \mathrm{otherwise}%
\end{array}%
\right. ~.  \label{rho}
\end{equation}

Chau and Yue \cite{Chau Yu} have subsequently introduced ensembles of random
normal matrices in the context of the quantum Hall problem of $N$ electrons
in a strong magnetic field, opening a new front of research in the area of
random matrices. Since normal matrices are unitarily equivalent to a
diagonal matrix, the probability distribution of eigenvalues for random
normal ensembles can be achieved, exactly as in the Hermitian ensembles, by
choosing an appropriated coordinate system that factorizes the eigenvalues
contribution from the rest (see, respectively, Section $5.3$ of \cite{Deift}
and \cite{Chau Zaboronsky} for the Hermitian and normal ensembles).

Normal ensembles differ from the Hermitian counterpart by the statistical
dependence of matrix elements even for Gaussian ensembles and, most
importantly, by the fact that their eigenvalues are generically complex.
Among the usual questions concerning the statistics of their eigenvalues
there are some related with universality that remain unresolved for the
normal ensembles. According to the theory of random matrices, the eigenvalue
correlations in Hermitian, and normal ensembles as well, are given by the
determinant of an integral kernel whose asymptotic behavior for large $N$
governs their decay. The limit integral kernel is well known to be universal
for standard models of Hermitian ensembles (see \cite{Deift} and references
therein). The scenery for normal ensembles, despite of certain efforts in
this direction, remains undisclosed.

The present work addresses the integral kernel of ensembles of normal
matrices weighed by $e^{-NV}$ with $V$ depending only on the absolute value
of eigenvalues. We apply the steepest descents method to obtain scaling
limits for the integral kernel with error estimates in power of $1/N$. Our
results can be extended for a large class of radial symmetric potentials $V$
satisfying condition (\ref{1.3}) but we shall restrict ourselves to a sub
class of potentials (\ref{Va}), for simplicity. Although Chau and Zaboronsky 
\cite{Chau Zaboronsky} have given asymptotic expressions for one and
two--point correlation functions, the integral kernel of normal random
matrices has not been previously considered for the models addressed here.

The eigenvalue probability distribution of the ensemble of random normal
matrices is given by%
\begin{equation}
P_{N}(z_{1},\cdots ,z_{N})=Z_{N}^{-1}\emph{e}^{-N\sum_{i=1}^{N}V(z_{i})}%
\prod_{1\leq i<j\leq N}\left\vert z_{i}-z_{j}\right\vert ^{2}  \label{P1}
\end{equation}%
with potentials $V:\mathbb{C}\longrightarrow \mathbb{R}$ satisfying the
properties: $(i)$ $V$ is continuous and $(ii)$ 
\begin{equation}
\text{ \ }\underset{\left\vert z\right\vert \rightarrow \infty }{\lim }%
\left( \frac{V\left( z\right) }{2}-\log z\right) =\infty  \label{1.3}
\end{equation}%
to avoid the eigenvalues escape to infinity (see e.g. Saff and Totik \cite%
{Saff}). Equation (\ref{P1}) can be written as 
\begin{equation}
P_{N}(z_{1},\cdots ,z_{N})=\frac{1}{N!}\det \left( K_{N}\left(
z_{i},z_{j}\right) \right) _{i,j=1}^{N}  \label{P2}
\end{equation}%
with $K_{N}$ being the integral kernel 
\begin{equation}
K_{N}\left( z,w\right) =e^{-\frac{N}{2}V(z)}e^{-\frac{N}{2}\overline{V(w)}%
}\sum_{j=1}^{N}\phi _{j}\left( z\right) \overline{\phi _{j}\left( w\right) }
\label{KN}
\end{equation}%
where $\left\{ \phi _{j}\right\} _{j=1}^{N}$ is the set of the orthonormal
polynomials with respect to the inner product $\left( \cdot ,\cdot \right)
_{\nu _{N}}$ with weight 
\begin{equation*}
d\nu _{N}\left( z\right) =e^{-NV(z)}d^{2}z
\end{equation*}%
and the $n$--point correlation function associated to $P_{N}$ can be written
as%
\begin{equation}
R_{n}^{N}(z_{1},\cdots ,z_{n})=\det \left( K_{N}\left( z_{i},z_{j}\right)
\right) _{i,j=1}^{n}.  \label{Rn}
\end{equation}%
The statistics of the eigenvalues are thus given by the asymptotic behavior
of the integral kernel.

The main result of this paper is as follows.

\begin{theorem}
\label{akna} Let 
\begin{equation}
V_{\alpha }\left( z\right) =\left\vert z\right\vert ^{\alpha }\ ,\ \alpha >0
\label{Va}
\end{equation}%
be a family of radially symmetric potentials, 
\begin{equation*}
S(\tau ,K)=\left\{ \zeta \in \mathbb{C}:0<\left\vert \zeta \right\vert <K~,\
\left\vert \arg \zeta \right\vert <\frac{\tau }{2}\right\}
\end{equation*}%
be a sectorial domain of opening $\tau $ and radius $K$ and, for each $%
0<\delta <1$, let $\gamma =\gamma (\alpha ,\delta )$ be such that 
\begin{equation*}
\alpha \gamma +\delta =1~.
\end{equation*}%
Then, the integral kernel (\ref{KN}) with $V=V_{\alpha }$ satisfies 
\begin{equation}
\frac{1}{N^{\delta +2\gamma }}K_{N}^{\alpha }\left( \frac{Z}{N^{\gamma }},%
\frac{W}{N^{\gamma }}\right) =\frac{\alpha ^{2}}{4\pi }\left( Z\bar{W}%
\right) ^{\frac{\alpha }{2}-1}e^{N^{\delta }\left( \left( Z\bar{W}\right) ^{%
\frac{\alpha }{2}}-\frac{\left\vert Z\right\vert ^{\alpha }}{2}-\frac{%
\left\vert W\right\vert ^{\alpha }}{2}\right) }\left( 1+E_{N}^{\alpha
,\delta }(Z\bar{W})\right)  \label{knazwdc}
\end{equation}%
with error estimation 
\begin{equation}
\left\vert E_{N}^{\alpha ,\delta }(\zeta )\right\vert \leq O\left(
N^{-\delta /2}\right)  \label{E}
\end{equation}%
whenever $\zeta \in S\left( \theta /\sqrt{N},(2/\alpha )^{2/\alpha
}N^{2(1-\delta )/\alpha }\right) $, for some $\theta >0$ and large enough $N$%
.

In particular, taking $\delta \nearrow 1$ and, consequently, $\gamma
\searrow 0$ we obtain 
\begin{equation}
\frac{1}{N}K_{N}^{\alpha }\left( Z,W\right) =\frac{\alpha ^{2}}{4\pi }\left(
Z\bar{W}\right) ^{\frac{\alpha }{2}-1}e^{N\left( \left( Z\bar{W}\right) ^{%
\frac{\alpha }{2}}-\frac{\left\vert Z\right\vert ^{\alpha }}{2}-\frac{%
\left\vert W\right\vert ^{\alpha }}{2}\right) }\left( 1+E_{N}^{\alpha ,1}(Z%
\bar{W})\right)  \label{kd1}
\end{equation}%
with $O\left( 1/\sqrt{N}\right) $ error for $Z\bar{W}\in S\left( \theta /%
\sqrt{N},(2/\alpha )^{2/\alpha }\right) $.
\end{theorem}

\begin{remark}
The parameter $\delta <1$ has been introduced to ensure that the eigenvalues
are "sampling" in the bulk, out of any fixed compact domain containing the
origin. The case of interest for applications is the limit point $\delta =1$%
. The limit, as $N$ goes to infinity, of any function involving the
asymptotic expression (\ref{kd1}) is called \textbf{bulk scaling limit} of
that function.
\end{remark}

\begin{remark}
The restriction to a sector $S(\tau ,K)$ of opening $\tau $ that shrinks
with $1/\sqrt{N}$ is an artifact of our method. Equation (\ref{knazwdc}) is
expected to hold for $Z\bar{W}\in S(\tau ,K)$, with $K=K(\tau )>0$ for $%
0\leq \tau <4\pi /\alpha $, but our estimates on the error for replacing a
sum by an integral, giving by the Euler--Maclaurin sum formula, break down
except for sectors $S(\theta N^{-\beta },K)$ with $\theta >0$ and $\beta
\geq 1/2$ (see (\ref{S1}) and following equations). Numerical calculations
performed in \cite{VPM1} for $\alpha \geq 2$ indicate that (\ref{knazwdc})
might hold for $Z\bar{W}\in S(4\pi /\alpha ,K)$ with an error decaying
faster than any power of $N$ for some $K<1$ (see also the next remark for an
improved and simple estimate for $\alpha =2$). There, a different error:%
\begin{equation*}
\sup_{\left\vert z\right\vert ,\left\vert w\right\vert <(2/\alpha
)^{1/\alpha };\left\vert \arg (z\bar{w})\right\vert <2\pi /\alpha
}\left\vert \frac{\alpha ^{2}}{4\pi }\left( Z\bar{W}\right) ^{\frac{\alpha }{%
2}-1}e^{N^{\delta }\left( \left( Z\bar{W}\right) ^{\frac{\alpha }{2}}-\frac{%
\left\vert Z\right\vert ^{\alpha }}{2}-\frac{\left\vert W\right\vert
^{\alpha }}{2}\right) }E_{N}^{\alpha ,1}(z\bar{w})\right\vert
\end{equation*}%
denoted by $R_{n}^{\alpha }$, has been considered. We warn that the result
of Theorem \ref{akna} has been imprecisely stated in Eq. (9) of \cite{VPM1}.
\end{remark}

\begin{remark}
Taylor remainder formula can be used to estimate the difference between the
Taylor polynomial $S_{N}$ and the function $f_{N}$, respectively defined by (%
\ref{KtildeNzw}) with $\delta =1$ (see also (\ref{soma})) and by the
infinite sum with the same summand. For $\alpha =2$, $f_{N}(\zeta )=N\zeta
e^{N\zeta }/\pi $. By (\ref{KNzw}), together with the Lagrange remainder,
one gets (\ref{kd1}) with the error function satisfying $\left\vert
E_{N}(\zeta )\right\vert =O\left( N^{-1/2}(e\left\vert \zeta \right\vert
)^{N}e^{-N(1-a)\Re \mathrm{e}\zeta }\right) $, for some $0<a<1$ and large
enough $N$ (see calculations in Appendix \ref{TR}). We observe that (\ref%
{kd1}) with $\alpha =2$ holds with $\sup_{\zeta \in \bar{S}(\tau
,K)}\left\vert E_{N}(\zeta )\right\vert =O(1/\sqrt{N})$ for $\zeta =Z\bar{W}$
in a sectorial domain $S(\tau ,K)$ with $K=K(\tau ,a)>0$ given by smallest
solution of $Ke^{-(1-a)K\cos \tau /2+1}=1$.\footnote{%
We thank an anonymous referee for suggesting the use of the Taylor remainder
to estimate the error in (\ref{kd1}) for $\alpha =2$.} Taylor remainder
method, together (perhaps) with some additional ingredient, may be extended
for $\alpha >2$ but it doesn't seems to work for $0<\alpha <2$.
\end{remark}

\begin{remark}
The asymptotic behavior (\ref{kd1}) for $\alpha =2$, without error
estimation, was established in \cite{Fyodorov1}. Whether $n$--point
functions are universal for normal ensembles with weight $e^{-NV}$, where $%
V(z)$ is a polynomial in $\left\vert z\right\vert ^{2}$ of positive degree
with nonnegative coefficients, was addressed in \cite{Chau Zaboronsky}.
\end{remark}

It follows from equations (\ref{kd1}) and (\ref{Rn}) that normal ensembles
with the class of potentials $V_{\alpha }$ are universal alike the Hermitian
ensembles (see e.g. Subsection $5.6.1$ of \cite{Deift} and \cite{LL}, for
recent results):

\begin{corollary}
\label{npoint}Let $r,z_{1},\ldots ,z_{n}$ be $n+1$ complex numbers and write%
\footnote{%
For Hermitian ensembles, $z_{i}/\sqrt{\pi K_{N}^{\alpha }(r,r)}$ and the
universal integral kernel $\mathbb{K}(z,w)$ in (\ref{R}) are respectively
replaced by $x_{i}/(\pi K_{N}^{\alpha }(r,r))$ and by the Sinc function $%
\mathbb{S}\left( x-y\right) =\dfrac{\sin (x-y)}{\pi (x-y)}$.}%
\begin{equation}
Z_{i}=r+\frac{z_{i}}{\sqrt{\pi K_{N}^{\alpha }(r,r)}}~.  \label{Zi}
\end{equation}%
Then, the following scaling limit for the $n$--point function 
\begin{equation}
\lim_{N\rightarrow \infty }\frac{1}{\pi ^{n}K_{N}^{\alpha }(r,r)^{n}}%
R_{n}^{N}\left( Z_{1},\ldots ,Z_{n}\right) =\det \left( \mathbb{K}\left(
z_{i},z_{j}\right) \right) _{i,j=1}^{n}  \label{R}
\end{equation}%
holds uniformly for $r$ in any compact set of the open set $\left\{ z\in 
\mathbb{C}:0<\left\vert z\right\vert <(2/\alpha )^{1/\alpha }\right\} $,
where 
\begin{equation}
\mathbb{K}(z,w)=\dfrac{1}{\pi }e^{\left( z\bar{w}-\frac{\left\vert
z\right\vert ^{2}}{2}-\frac{\left\vert w\right\vert ^{2}}{2}\right) }
\label{K}
\end{equation}%
is the integral kernel for the classical Segal--Bargmann space of entire
functions. The bulk scaling limit (\ref{R}) is universal in the sense that
it is independent of the family of potentials $V_{a}$.
\end{corollary}

We shall address this and other issues related with the conformal invariance
of the integral kernel (\ref{KN}) in a forthcoming paper \cite{VPM}. Since
the cancellations involved makes the implication of (\ref{R}) far of being
straightforward, a complete, although short, proof has been included in
Appendix \ref{PCnpoint}.

For $n=2$, (\ref{R}) reads%
\begin{eqnarray}
\lim_{N\rightarrow \infty }\frac{1}{\pi ^{2}K_{N}^{\alpha }(r,r)^{2}}%
R_{n}^{N}\left( Z_{1},Z_{2}\right) &=&\left( \mathbb{K}(z_{1},z_{1})\mathbb{K%
}(z_{2},z_{2})-\mathbb{K}(z_{1},z_{2})\mathbb{K}(z_{2},z_{1})\right)  \notag
\\
&=&\frac{1}{\pi ^{2}}\left( 1-e^{-\left\vert z_{1}-z_{2}\right\vert
^{2}}\right) ~,  \label{2point}
\end{eqnarray}%
a result already obtained for more general radial potentials (see Theorem 1
of \cite{Chau Zaboronsky}). Under the assumption that (\ref{kd1}) can be
extended to the sectorial domain $S(4\pi /\alpha ,K)$ (this actually holds
for $\alpha =2$. See Appendix \ref{TR}), a change of variables in the
integral Kernel by the function $\varphi _{N}(z)=(z/\sqrt{N})^{2/\alpha }$,
which maps conformally $\left\{ z\in \mathbb{C}:\left\vert z\right\vert
<K^{\alpha /2}\sqrt{N}\right\} $ into $S(4\pi /\alpha ,K)$, yields%
\begin{equation}
\lim_{N\rightarrow \infty }\varphi _{N}^{\prime }\left( z\right)
K_{N}^{\alpha }\left( \varphi _{N}\left( z\right) ,\varphi _{N}\left(
w\right) \right) \overline{\varphi _{N}^{\prime }\left( w\right) }=\mathbb{K}%
(z,w)~  \label{conform}
\end{equation}%
where $\mathbb{K}(z,w)$ is the integral kernel given by (\ref{K}). This
notion of universality has been called conformal universality in \cite{VPM1}%
. The estimates in Appendix \ref{TR} establishes the pointwise limit (\ref%
{conform}) in $\mathbb{C}\times \mathbb{C}$ for $\alpha =2$.

Theorem \ref{akna} will be proven in Section \ref{PTakna}. Sections \ref{RNE}
and \ref{IK} contain preliminary materials. The technical part of our result
concerns with the error estimation of Euler--Maclaurin formula. Different
methods needs to be employed depending on the regions considered in the sum.
Appendix \ref{TR} estimates the Taylor remainder of (\ref{KtildeNzw}) for $%
\delta =1$ and $\alpha =2$ and Appendix \ref{PCnpoint} proves Corollary \ref%
{npoint}.

\section{Ensemble of random normal matrices \textit{\label{RNE}}}

\setcounter{equation}{0} \setcounter{theorem}{0}

We begin with the following

\begin{definition}
By normal ensembles we mean a probability measure 
\begin{equation}
P(M_{N})dM_{N}=Z_{N}^{-1}e^{-N\text{Tr}V(M_{N})}dM_{N}  \label{P}
\end{equation}%
on the set of $N\times N$ complex matrices $M_{N}$ supported on the variety $%
\left[ M_{N},M_{N}^{\ast }\right] =0$ and invariant by unitary conjugation $%
\tilde{M}_{N}=U_{N}^{\ast }M_{N}U_{N}$: 
\begin{equation}
P\left( M_{N}\right) dM_{N}=P\left( \tilde{M}_{N}\right) d\tilde{M}_{N}~.
\label{PMn}
\end{equation}
\end{definition}

The elements $m_{ij}=m_{ij}^{R}+im_{ij}^{I}$, $1\leq i\leq j\leq N$ of $%
M_{N} $ in the normal ensemble cannot be picked independently according to
any product measure, absolutely continuous with respect to the Lebesgue
measure $\displaystyle\prod_{1\leq i\leq j\leq N}dm_{ij}^{R}dm_{ij}^{I}$ in $%
\mathbb{R}^{N^{2}+N}$, even when the weight $e^{-N\text{Tr}V(M_{N})}$ is
Gaussian, in view of the constraint on elements $m_{ij}$ with $i>j$ 
\footnote{%
If they were independent, it would contradict Schur--Toeplitz statement (see
e.g. \cite{Lancaster}): \textquotedblleft any square matrix is unitarily
similar to an upper (or lower) triangular matrix\textquotedblright .}. So,
the elements of $M_{N}$ when sampling on normal ensembles are always
statistically dependent. Note that the set of normal matrices with simple
spectra is open and dense in $\mathbb{R}^{N^{2}+N}$ and has full measure
(see \cite{Deift} for a proof in the Hermitian ensembles).

As $M_{N}$ is normal, $M_{N}$ is unitarily equivalent to a diagonal matrix
of eigenvalues and there exist $U_{N}$ satisfying $U_{N}^{-1}=$ $U_{N}^{\ast
}$ and%
\begin{equation}
M_{N}=U_{N}\Lambda _{N}U_{N}^{\ast }  \label{diagonaliza}
\end{equation}%
with $\Lambda _{N}=\mathrm{diag}\left\{ z_{1},\cdots ,z_{N}\right\} $,
ordered according their absolute value: $\left\vert z_{i}\right\vert \leq
\left\vert z_{j}\right\vert $ if $i<j$. Following section $5.3$ of \cite%
{Deift} with few adjustments (see \cite{Chau Zaboronsky} and \cite%
{ElbauFelder}), the spectral decomposition (\ref{diagonaliza}) considered as
a change of variables $M_{N}\overset{\varphi }{\longmapsto }\left( \Lambda
_{N},U_{N}~\text{mod~}\mathbb{T}^{N}\right) $ yields 
\begin{equation}
P(M_{N})dM_{N}=Z^{-1}\emph{e}^{-N\sum_{i=1}^{N}V(z_{i})}J\left( z,p\right)
\prod_{1\leq i\leq N}d^{2}z_{i}\prod_{1\leq j\leq l}d^{2}p_{k},\text{ }
\label{Pex}
\end{equation}%
where $\{p_{i}\}_{i=1}^{l}$ with $2l+N=N^{2}$, are variables associated with
the eigenvectors of $M$, $d^{2}z$ denotes the Lebesgue measure in $\mathbb{C}
$ and%
\begin{equation*}
J\left( z,p\right) =\prod_{1\leq i<j\leq N}\left\vert z_{i}-z_{j}\right\vert
^{2}f(p)
\end{equation*}%
is the Jacobian of $\varphi $, with $f$ a function depending only on the
eigenvectors variables $\{p_{i}\}_{i=1}^{l}$. The eigenvalue probability
distribution (\ref{P}) of this ensemble is obtained integrating (\ref{Pex})
with respect to $\{p_{i}\}_{i=1}^{l}.$

The $n$-point correlation function is defined by (see e.g. \cite{Ruelle}) 
\begin{equation}
R_{n}^{N}(z_{1},\cdots ,z_{n})=\frac{N!}{\left( N-n\right) !}\int
P_{N}(z_{1},\cdots ,z_{N})\prod_{i=n+1}^{N}d^{2}z_{i}  \label{RnN}
\end{equation}%
and it can be written as (\ref{Rn}). Stochastic processes of this form are
called random determinantal point fields \cite{Soshnikov}. The present work
concerns with the asymptotic analysis of the integral kernel (\ref{KN}) and
its implications to the limit of the $n$-point correlation function. We have
seen that the limit of the $2$--point correlation (\ref{2point}) can be read
directly from the asymptotic formula (\ref{kd1}). The eigenvalue density $%
\rho ^{V_{\alpha }}$, associated with the normal ensemble defined by $%
V_{\alpha }$, is given by%
\begin{equation}
\rho ^{V_{\alpha }}\left( z\right) ={\lim_{N\rightarrow \infty }}\frac{1}{N}%
R_{1}^{N}(z)=\lim_{N\rightarrow \infty }\frac{1}{N}K_{N}\left( z,z\right) =%
\frac{\alpha ^{2}}{4\pi }\left\vert z\right\vert ^{\alpha -2}~  \label{ro}
\end{equation}%
for $\left\vert z\right\vert \leq \left( 2/\alpha \right) ^{1/\alpha }$ (see
Remarks \ref{bound}, for more comment on this). Note that $\rho
^{V_{a}}(z)d^{2}z$ and the equilibrium or extremal measure $d\hat{\sigma}(z)$
(see e.g. \cite{Hedenmalm}) agree and are supported on the same domain.

\section{Integral kernel of normal \textit{ensembles} defined by $V_{\protect%
\alpha }$ and various estimates\textit{\label{IK}}}

\setcounter{equation}{0} \setcounter{theorem}{0}

The present section is devoted to preliminary results on the integral kernel
(\ref{KN}).

Let $L^{2}\left( \mathbb{C},\nu \right) $ denote the Hilbert space of
square--integrable complex--valued functions 
\begin{equation*}
\left\Vert f\right\Vert _{\nu }^{2}=\int_{\mathbb{C}}\left\vert
f(z)\right\vert ^{2}d\nu (z)<\infty
\end{equation*}%
with respect to a positive finite Borel measure $\nu $ on $\mathbb{C}$
which, in order to ensure that all analytic polynomials belong to the space
is assumed to satisfy%
\begin{equation*}
\int_{\mathbb{C}}\left\vert z\right\vert ^{2n}d\nu (z)<\infty \ ,\ \ n\in 
\mathbb{N}\ .
\end{equation*}%
If $P_{N}\left( \mathbb{C},\nu \right) $ denotes the $N$--dimensional linear
vector space of analytic polynomials of degree less than or equal $N-1$,
endowed with the inner product%
\begin{equation}
\left( p,q\right) _{\nu }=\int_{\mathbb{C}}\overline{p(z)}q(z)d\nu (z)~,
\label{inner}
\end{equation}%
we have

\begin{proposition}
\label{cva}For each $N\in \mathbb{N}$, the monomials 
\begin{equation*}
\phi _{j}^{\alpha }\left( z\right) =\sqrt{\frac{\alpha }{2\pi \Gamma \left(
2j/\alpha \right) }}N^{j/\alpha }z^{j-1}
\end{equation*}%
with $j=1,\ldots ,N$, form an orthonormal set in $P_{N}\left( \mathbb{C},\nu
_{N}^{\alpha }\right) $ with respect to 
\begin{equation*}
d\nu _{N}^{\alpha }\left( z\right) =e^{-N\left\vert z\right\vert ^{\alpha
}}d^{2}z~,\ \alpha >0~.
\end{equation*}%
The integral kernel (\ref{KN}) reads in this case 
\begin{equation}
K_{N}^{\alpha }\left( z,w\right) =e^{-\frac{N}{2}\left\vert z\right\vert
^{\alpha }}e^{-\frac{N}{2}\left\vert w\right\vert ^{\alpha }}\tilde{K}%
_{N}^{\alpha }\left( z,w\right)  \label{KNzw}
\end{equation}%
where 
\begin{equation}
\tilde{K}_{N}^{\alpha }\left( z,w\right) =\frac{\alpha }{2\pi }\sum_{j=1}^{N}%
\frac{N^{2j/\alpha }\left( z\overline{w}\right) ^{j-1}}{\Gamma \left(
2j/\alpha \right) }  \label{KtildeNzw}
\end{equation}%
is a reproducing kernel on $P_{N}\left( \mathbb{C},\nu _{N}^{\alpha }\right) 
$.
\end{proposition}

\begin{remark}
For the Bergman space $A^{2}\left( \Omega \right) $ of square--integrable
single--valued analytic function on a compact domain $\Omega $ , there
always exist a complete set of orthonormal polynomials $\left\{ \phi
_{j}(z)\right\} _{j=1}^{\infty }$ and the integral kernel 
\begin{equation*}
\tilde{K}(z,w)=\sum_{j=1}^{\infty }\phi _{j}(z)\overline{\phi _{j}(w)}
\end{equation*}%
converges $\lim_{N\rightarrow \infty }\displaystyle\sum_{j=1}^{N}\phi _{j}(z)%
\overline{\phi _{j}(w)}=\tilde{K}(z,w)$ uniformly for any $z,w$ in $\Omega $ 
\cite{Bergman}. This is not necessarily the case for unbounded domain but
the same properties hold for Segal--Bargmann spaces $A^{2}\left( \mathbb{C}%
;\nu \right) $ of single--valued analytic functions in $\mathbb{C}$,
square--integrable with respect to $e^{-\left\vert z\right\vert ^{2}}d^{2}z$%
. We call the reader's attention to the $N$ dependence on the inner product (%
\ref{inner}) and the fact that the limit $N$ to infinity in (\ref{KNzw})
involves also a limit of the measure $\nu _{N}^{\alpha }$. As one sees from (%
\ref{kd1}), together with%
\begin{equation*}
\frac{\left\vert z\right\vert ^{\alpha }}{2}+\frac{\left\vert w\right\vert
^{\alpha }}{2}-\Re \mathrm{e}\left( z\bar{w}\right) ^{\alpha /2}=\frac{1}{2}%
\left\vert z^{\alpha /2}-w^{\alpha /2}\right\vert ^{2}\geq 0~,
\end{equation*}%
(equality iff $z=w$) and equation (\ref{ro}), the limit as $N\rightarrow
\infty $ of $K_{N}^{\alpha }\left( z,w\right) $ goes $0$ for $z\neq w$ and
diverges for $z=w$.
\end{remark}

We shall use (\ref{KtildeNzw}) to obtain an asymptotic expression as stated
in Theorem \ref{akna}.

\medskip

\noindent \textbf{Proof of Proposition \ref{cva}.} We need to verify that
the monomials are orthogonal with respect to the inner product (\ref{inner}%
). Writing 
\begin{equation*}
\phi _{j}\left( z\right) =\frac{z^{j-1}}{\sqrt{2\pi I_{j}}}
\end{equation*}%
with $z=re^{i\theta }$, we have%
\begin{eqnarray*}
\left( \phi _{k}\left( z\right) ,\phi _{j}\left( z\right) \right) _{\nu
_{N}^{\alpha }} &=&\frac{1}{2\pi \sqrt{I_{k}I_{j}}}\int \overline{z^{k}}%
z^{j}e^{-N\left\vert z\right\vert ^{\alpha }(z)}d^{2}z \\
&=&\frac{1}{\sqrt{I_{k}I_{j}}}\int_{0}^{\infty }r^{k+j+1}e^{-Nr^{\alpha }}dr%
\frac{1}{2\pi }\int_{0}^{2\pi }e^{i\theta \left( j-k\right) }d\theta =\delta
_{k,j}~,
\end{eqnarray*}%
with the Kroneker delta function $\delta _{k,j}=1$ if $k=j$ and $0$
otherwise, provided%
\begin{equation*}
I_{j}=\int_{0}^{\infty }r^{2j-1}e^{-Nr^{\alpha }}dr=\frac{N^{-2j/\alpha }}{%
\alpha }\Gamma \left( \frac{2j}{\alpha }\right) ~.
\end{equation*}%
Consequently, any analytic polynomial $p(z)$ in $P_{N}\left( \mathbb{C},\nu
_{N}^{\alpha }\right) $ can be written as 
\begin{equation}
p(z)=\sum_{j=1}^{N}c_{j}\phi _{j}\left( z\right)  \label{series}
\end{equation}%
with Fourier coefficients 
\begin{equation}
c_{j}=\left( \phi _{j},p\right) _{\nu _{N}^{\alpha }}=\int_{\mathbb{C}}%
\overline{\phi _{j}(w)}p(w)e^{-N\left\vert w\right\vert ^{\alpha }}d^{2}w~.
\label{cj}
\end{equation}%
Inserting (\ref{cj}) into (\ref{series}), gives $p(z)=\left( \overline{%
\tilde{K}_{N}^{\alpha }(z,\cdot )},p\right) _{\nu ^{\alpha }}$ where%
\begin{equation}
\tilde{K}_{N}^{\alpha }\left( z,w\right) =\sum_{j=1}^{N}\phi _{j}\left(
z\right) \overline{\phi _{j}\left( w\right) }=\frac{\alpha }{2\pi }%
\sum_{j=1}^{N}\frac{N^{2j/\alpha }\left( z\overline{w}\right) ^{j-1}}{\Gamma
\left( 2j/\alpha \right) }~.  \label{Ktilde}
\end{equation}

\hfill $\Box $

Looking for an asymptotic expansion of (\ref{KNzw}), a complex valued
function is defined on the positive real line $\mathbb{R}_{+}=\left(
0,\infty \right) $ coinciding with the summand of the integral kernel (\ref%
{Ktilde}) on $\mathbb{N}$. For fixed numbers $\alpha >0$, $0<\delta <1$, $%
\zeta \in \mathbb{C}\backslash \left\{ 0\right\} $ and $N$ a positive
integer, let $g_{\zeta }:\mathbb{R}_{+}\longrightarrow \mathbb{C}$ be given
by 
\begin{equation}
g_{\zeta }\left( x\right) =\frac{\left( N^{\frac{2\delta }{\alpha }}\zeta
\right) ^{x}}{\Gamma \left( 2x/\alpha \right) }  \label{ge}
\end{equation}%
and note that $\left\vert g_{\zeta }(x)\right\vert =g_{\left\vert \zeta
\right\vert }(x)$.

\begin{lemma}
\label{bor}Under the above conditions on $\alpha $, $\delta $, $\zeta $ and $%
N$, the real valued function $g_{\left\vert \zeta \right\vert }:\mathbb{R}%
_{+}\longrightarrow \mathbb{R}$ has a global maximum 
\begin{equation*}
g_{\left\vert \zeta \right\vert }(x)\leq \max_{x\geq 0}g_{\left\vert \zeta
\right\vert }(x)=g_{\left\vert \zeta \right\vert }(x^{\ast })
\end{equation*}%
at $x^{\ast }=x^{\ast }(\alpha ,\delta ,\left\vert \zeta \right\vert ,N)>0$.
For $N$ large enough so that $N>N_{0}$, 
\begin{equation}
N_{0}=\max \left( \left( \frac{k}{\left\vert \zeta \right\vert }\right) ^{%
\frac{\alpha }{2\delta }},\left( \frac{\alpha }{2}\left\vert \zeta
\right\vert ^{\frac{\alpha }{2}}\right) ^{\frac{1}{1-\delta }}\right)
\label{N>}
\end{equation}%
with $k$\ a large universal constant, the inequality 
\begin{equation*}
0<x^{\ast }<N
\end{equation*}%
holds and%
\begin{eqnarray}
g_{\left\vert \zeta \right\vert }(x^{\ast }) &=&\frac{1}{\sqrt{2\pi }}%
\left\vert \zeta \right\vert ^{\frac{\alpha }{4}}N^{\frac{\delta }{2}}\exp
\left( \left\vert \zeta \right\vert ^{\frac{\alpha }{2}}N^{\delta }\right)
\left( 1+O\left( \frac{1}{N^{\delta }}\right) \right)  \label{gm} \\
x^{\ast } &=&\frac{\alpha }{2}\left\vert \zeta \right\vert ^{\frac{\alpha }{2%
}}N^{\delta }-\frac{\alpha }{4}+O\left( \frac{1}{N^{\delta }}\right)
\label{m}
\end{eqnarray}
\end{lemma}

\noindent \textbf{Proof.} Differentiating $g_{\left\vert \zeta \right\vert
}(x)$ with respect to $x$, we have 
\begin{equation}
g_{\left\vert \zeta \right\vert }^{\prime }(x)=g_{\left\vert \zeta
\right\vert }\left( x\right) \left( \log \left( N^{\frac{2\delta }{\alpha }%
}\left\vert \zeta \right\vert \right) -\frac{2}{\alpha }\psi \left( \frac{2}{%
\alpha }x\right) \right)  \label{glinha}
\end{equation}%
where $\psi \left( x\right) =\Gamma ^{\prime }\left( x\right) /\Gamma \left(
x\right) $ is the digamma function. Since $g_{\left\vert \zeta \right\vert
}\left( x\right) $ does not vanish and $\psi \left( x\right) $ belongs to a
Pick class of functions that can be analytically continued through $\mathbb{R%
}_{+}$ (see e.g. \cite{Donoghue}), as $x$ varies in the semi--line $\psi
\left( x\right) $ increases monotonously from $-\infty $ to $\infty $ and
the maximum of $g_{\left\vert \zeta \right\vert }$ is attained at the unique
solution $x=x^{\ast }$ of 
\begin{equation}
\log \left( N^{\frac{2\delta }{\alpha }}\left\vert \zeta \right\vert \right)
-\frac{2}{\alpha }\psi \left( \frac{2}{\alpha }x\right) =0.  \label{gl=0}
\end{equation}%
For $N$ so large that the asymptotic expansion \cite{Abramovitz} 
\begin{equation}
\psi \left( y\right) \sim \log y+\frac{1}{2y}-\sum_{j=1}^{\infty }B_{2j}%
\frac{1}{2jy^{2j}}  \label{digamma}
\end{equation}%
of digamma function at $y=N^{\frac{2\delta }{\alpha }}\left\vert \zeta
\right\vert $ can be applied (i. e., $y>k$ where $k$ is the constant mention
in (\ref{N>})), we have by (\ref{gl=0}) 
\begin{equation*}
\log \left( N^{\delta }\left\vert \zeta \right\vert ^{\frac{\alpha }{2}%
}\right) =\log \frac{2}{\alpha }x^{\ast }+\frac{\alpha }{4x^{\ast }}+O\left( 
\frac{1}{x^{\ast 2}}\right)
\end{equation*}%
or equivalently, 
\begin{equation*}
\frac{\alpha N^{\delta }\left\vert \zeta \right\vert ^{\frac{\alpha }{2}}}{2}%
=x^{\ast }+\frac{\alpha }{4}+O\left( \frac{1}{x^{\ast }}\right)
\end{equation*}%
which establishes (\ref{m}). The coefficients $B_{2j}$ in (\ref{digamma})
are the Bernoulli numbers:%
\begin{equation*}
\frac{t}{e^{t}-1}=\sum_{n=0}^{\infty }B_{n}\frac{t^{n}}{n!}~.
\end{equation*}

For (\ref{N>}), it suffices to solve $\alpha N^{\delta }\left\vert \zeta
\right\vert ^{\frac{\alpha }{2}}/2\leq N$ for $N.$ For (\ref{gm}), we plug (%
\ref{m}) into $g_{\left\vert \zeta \right\vert }(x^{\ast })$. As $x^{\ast }$
is order $N^{\delta }$ and, therefore, large enough for applying Stirling
formula, 
\begin{eqnarray}
g_{\left\vert \zeta \right\vert }(x^{\ast }) &=&\frac{\left( N^{\frac{%
2\delta }{\alpha }}\left\vert \zeta \right\vert \right) ^{x^{\ast }}}{\Gamma
\left( \frac{2}{\alpha }x^{\ast }\right) }  \notag \\
&=&\sqrt{\frac{x^{\ast }}{\alpha \pi }}\left( \frac{\alpha e}{2x^{\ast }}%
\right) ^{\frac{2}{\alpha }x^{\ast }}\left( \left\vert \zeta \right\vert N^{%
\frac{2\delta }{\alpha }}\right) ^{x^{\ast }}\left( 1+O\left( \frac{1}{%
N^{\delta }}\right) \right)  \notag \\
&=&\frac{\left\vert \zeta \right\vert ^{\frac{\alpha }{4}}}{\sqrt{2\pi }}N^{%
\frac{\delta }{2}}e^{N^{\delta }\left\vert \zeta \right\vert ^{\frac{\alpha 
}{2}}}\left( 1+O\left( \frac{1}{N^{\delta }}\right) \right) ~.  \label{gmx}
\end{eqnarray}

\hfill $\Box $

\begin{remark}
\label{bound}Lemma \ref{bor} still holds for $\delta =1$ provided $%
0<\left\vert \zeta \right\vert \leq (2/\alpha )^{2/\alpha }$. Note that $%
x^{\ast }=N-\alpha /4+O\left( 1/N\right) <N$ for $\left\vert \zeta
\right\vert =(2/\alpha )^{2/\alpha }$, which defines the domain boundary of
the density of eigenvalues (\ref{ro}) (recall $\zeta =Z\bar{W}$ and $%
\left\vert Z\right\vert ,\left\vert W\right\vert \leq (2/\alpha )^{1/\alpha
} $).
\end{remark}

The limit $\lim_{N\rightarrow \infty }\tilde{K}_{N}^{\alpha }(z,w)/N$
calculated at $z\bar{w}=\zeta /N^{2/\alpha }$, given by the series 
\begin{equation*}
\left( \alpha /2\pi \right) \sum_{j=1}^{\infty }\zeta ^{j-1}/\Gamma \left(
2j/\alpha \right) \;,
\end{equation*}%
converges uniformly in compact sets of $\mathbb{C}$ to an entire function of 
$\zeta $ of order $\alpha /2$, whose maximum is determined, essentially, by
a single term of the series, the so called central index $j^{\ast }=j^{\ast
}(\left\vert \zeta \right\vert )$. The next result estimates the range of
indices $j$ in (\ref{Ktilde}) the contributes for its asymptotic expansion
for large $N$.

\begin{lemma}
\label{distmax}Let $x$ be a point that is at least $N^{\frac{\delta }{2}%
}\log N$ away from the global maximum (\ref{m}) of $g_{\left\vert \zeta
\right\vert }\left( x\right) $, that is, 
\begin{equation}
\left\vert x-x^{\ast }\right\vert \geq N^{\frac{\delta }{2}}\log N.
\label{xx}
\end{equation}%
Then 
\begin{equation}
g_{\left\vert \zeta \right\vert }\left( x\right) \leq \max \left(
g_{\left\vert \zeta \right\vert }\left( x_{+}\right) ,g_{\left\vert \zeta
\right\vert }\left( x_{-}\right) \right)  \label{gx+-}
\end{equation}%
where $x_{\pm }=x^{\ast }\pm N^{\frac{\delta }{2}}\log N$ and%
\begin{equation}
g_{\left\vert \zeta \right\vert }\left( x_{\pm }\right) =\frac{1}{N^{2\log
N/(\alpha ^{2}\left\vert \zeta \right\vert ^{\alpha /2})}}g_{\left\vert
\zeta \right\vert }\left( x^{\ast }\right) \left( 1+O\left( \frac{\log ^{3}N%
}{N^{\delta /2}}\right) \right) ~.  \label{g+-}
\end{equation}
\end{lemma}

\noindent \textbf{Proof.} (\ref{gx+-}) follows by uniqueness of the maximum
value. For (\ref{g+-}), we repeat the estimates that lead to (\ref{gmx})
with $x_{\pm }$ in the place of $x^{\ast }$:%
\begin{equation}
g_{\left\vert \zeta \right\vert }\left( x_{\pm }\right) =\sqrt{\frac{x_{\pm }%
}{\alpha \pi }}\left( \frac{e\alpha N^{\delta }\left\vert \zeta \right\vert
^{\frac{\alpha }{2}}}{2x_{\pm }}\right) ^{2x_{\pm }/\alpha }\left( 1+O\left( 
\frac{1}{N^{\delta }}\right) \right) ~.  \label{gmelogn}
\end{equation}%
Plugging%
\begin{equation*}
x_{\pm }=\frac{\alpha }{2}\left\vert \zeta \right\vert ^{\frac{\alpha }{2}%
}N^{\delta }\pm N^{\frac{\delta }{2}}\log N-\frac{\alpha }{4}+O\left( \frac{1%
}{N^{\delta }}\right)
\end{equation*}%
into each term that appears in (\ref{gmelogn}), yields 
\begin{equation*}
\sqrt{\frac{x_{\pm }}{\alpha \pi }}=\sqrt{\frac{N^{\delta }\left\vert \zeta
\right\vert ^{\frac{\alpha }{2}}}{2\pi }}\left( 1+O\left( \frac{\log N}{N^{%
\frac{\delta }{2}}}\right) \right) ~,
\end{equation*}%
\begin{eqnarray*}
\frac{e\alpha N^{\delta }\left\vert \zeta \right\vert ^{\frac{\alpha }{2}}}{%
2x_{\pm }} &=&e\left( 1\pm \frac{2}{\alpha \left\vert \zeta \right\vert
^{\alpha /2}}\frac{\log N}{N^{\delta /2}}-\frac{1}{2\left\vert \zeta
\right\vert ^{\alpha /2}}\frac{1}{N^{\delta }}+O\left( \frac{1}{N^{2\delta }}%
\right) \right) ^{-1} \\
&=&\exp \left( 1\mp \frac{2}{\alpha \left\vert \zeta \right\vert ^{\alpha /2}%
}\frac{\log N}{N^{\delta /2}}+\frac{2}{\alpha ^{2}\left\vert \zeta
\right\vert ^{\alpha }}\frac{\log ^{2}N}{N^{\delta }}+\frac{1}{2\left\vert
\zeta \right\vert ^{\alpha /2}}\frac{1}{N^{\delta }}+O\left( \frac{\log N}{%
N^{3\delta /2}}\right) \right) ~,
\end{eqnarray*}%
where we have used%
\begin{equation*}
\frac{e}{1+\kappa }=\exp \left( 1-\log (1+\kappa )\right) =\exp \left(
1-\kappa +\frac{\kappa ^{2}}{2}+O\left( \kappa ^{3}\right) \right)
\end{equation*}%
and, therefore,%
\begin{equation*}
\left( \frac{e\alpha N^{\delta }\left\vert \zeta \right\vert ^{\frac{\alpha 
}{2}}}{2x_{\pm }}\right) ^{2x_{\pm }/\alpha }=\exp \left( \left\vert \zeta
\right\vert ^{\alpha /2}N^{\delta }-\frac{2}{\alpha ^{2}\left\vert \zeta
\right\vert ^{\alpha /2}}\log ^{2}N\right) \left( 1+O\left( \frac{\log ^{3}N%
}{N^{\delta /2}}\right) \right)
\end{equation*}%
Replacing in (\ref{gmelogn}), together with (\ref{gm}), results (\ref{g+-}).

\hfill $\Box $

We need one more ingredient.

\begin{lemma}
\label{Born}Let $f:\left[ a,b\right] \longrightarrow \mathbb{R}$ be a convex
function: 
\begin{equation*}
f\left( \lambda x+(1-\lambda )y\right) \leq \lambda f\left( x\right)
+(1-\lambda )f\left( y\right) \,
\end{equation*}%
for any $x,y\in \left[ a,b\right] $ and $0<\lambda <1$and let 
\begin{equation*}
P:a=x_{0}<\cdots <x_{K}=b
\end{equation*}%
be the partition of $\left[ a,b\right] $ into $K$ equally spacing
subintervals of length $\Delta $:\qquad\ 
\begin{equation*}
x_{j}=a+j\Delta ~,\ \ \ \ \,\,j\in \left\{ 0,\ldots ,K\right\} .
\end{equation*}%
Define $t_{j}\in \left[ x_{j},x_{j+1}\right] $ by the mean value theorem: 
\begin{equation}
\int_{x_{j}}^{x_{j+1}}f\left( x\right) dx=f\left( t_{j}\right) \Delta ~.
\label{k2j}
\end{equation}%
Then, the error in the trapezoidal approximation to the integral 
\begin{equation}
\Sigma (f;P):=\sum_{j=0}^{K-1}\left( \int_{x_{j}}^{x_{j+1}}f\left( x\right)
dx-\frac{1}{2}\left( f\left( x_{j}\right) +f\left( x_{j+1}\right) \right)
\Delta \right)  \label{ss}
\end{equation}%
is bounded by%
\begin{equation}
0\geq \Sigma (f;P)\geq \left( -\frac{f\left( t_{0}\right) }{2}+\frac{f\left(
x_{1}\right) }{2}+\frac{f\left( x_{K}\right) }{2}-\frac{f\left( t_{K}\right) 
}{2}\right) \Delta ~.  \label{som}
\end{equation}
\end{lemma}

\noindent \textbf{Proof.} Without loss of generality, we suppose that $f$ is
a positive convex function. Let $\left\{ k_{j}\right\} _{j=0}^{2K}$ be a
numerical sequence defined by 
\begin{eqnarray}
k_{2j} &=&\int_{x_{j}}^{x_{j+1}}f\left( x\right) dx  \notag \\
k_{2j+1} &=&f\left( x_{j+1}\right) \Delta \,  \label{kimpar}
\end{eqnarray}%
for $j\in \left\{ 0,\ldots ,K-1\right\} $ and note that, by the mean value
theorem (\ref{k2j}),%
\begin{equation}
k_{2j}=f\left( t_{j}\right) \Delta  \label{kpar}
\end{equation}%
for some $t_{j}\in \left[ x_{j},x_{j+1}\right] $. We shall prove, by a
geometric argument together with the convexity of $f$, that the following
inequality%
\begin{equation}
k_{i}\leq \frac{k_{i+1}+k_{i-1}}{2}\,  \label{med}
\end{equation}%
holds for each $i\in \left\{ 1,\ldots ,2K-1\right\} $.

Since $f$ is convex, the inequality (\ref{med}) for $i=2j$: 
\begin{equation*}
\int_{x_{j}}^{x_{j+1}}f\left( x\right) dx=k_{2j}\leq \frac{k_{2j+1}+k_{2j-1}%
}{2}=\frac{f\left( x_{j+1}\right) +f\left( x_{j}\right) }{2}\Delta
\end{equation*}%
is verified comparing the area under the function $f$ in the interval $\left[
x_{j},x_{j+1}\right] $ (left side of (\ref{med})) with the area of a
trapezoid formed by the points $\left( x_{j},0\right) ,$ $\left(
x_{j+1},0\right) ,$ $\left( x_{j},f\left( x_{j}\right) \right) $ and $\left(
x_{j+1},f\left( x_{j+1}\right) \right) $ (right side of (\ref{med})).

Once again, by convexity of $f$, the inequality (\ref{med}) for $i=2j+1$: 
\begin{equation}
f\left( x_{j+1}\right) \Delta =k_{2j+1}\leq \frac{k_{2j}+k_{2j+2}}{2}=\frac{1%
}{2}\,\int_{x_{j}}^{x_{j+2}}f\left( t\right) dt  \label{med1}
\end{equation}%
can be verified comparing the area under the function $f$ in the interval $%
\left[ x_{j},x_{j+2}\right] $ ($2\times $ the right side of (\ref{med1}))
with the area of a rectangle of base in the interval $\left[ x_{j},x_{j+2}%
\right] $ and height $f\left( x_{j+1}\right) $ ($2\times $ the left side of (%
\ref{med1})).

The later assertion is facilitate if the rectangle is replaced by a
trapezoid of same area obtained by rotating the horizontal segment at the
top around the point $\left( x_{j+1},f\left( x_{j+1}\right) \right) $ until
it becomes tangent to the graph of $f$ at that point (see figure below).

Now let us consider the sum 
\begin{eqnarray}
\Sigma _{1} &=&\sum_{j=0}^{2K}\left( -1\right) ^{j}k_{j}=k_{0}-k_{1}+\cdots
-k_{2K-1}+k_{2K}  \notag \\
&=&\frac{k_{0}}{2}-\sum_{j=0}^{K-1}\left( k_{2j+1}-\frac{k_{2j}+k_{2j+2}}{2}%
\right) +\frac{k_{2K}}{2}  \label{k1} \\
&=&k_{0}-\frac{k_{1}}{2}+\sum_{j=1}^{K-1}\left( k_{2j}-\frac{%
k_{2j-1}+k_{2j+1}}{2}\right) -\frac{k_{2K-1}}{2}+k_{2K}.  \label{k2}
\end{eqnarray}

\includegraphics[scale=0.4]{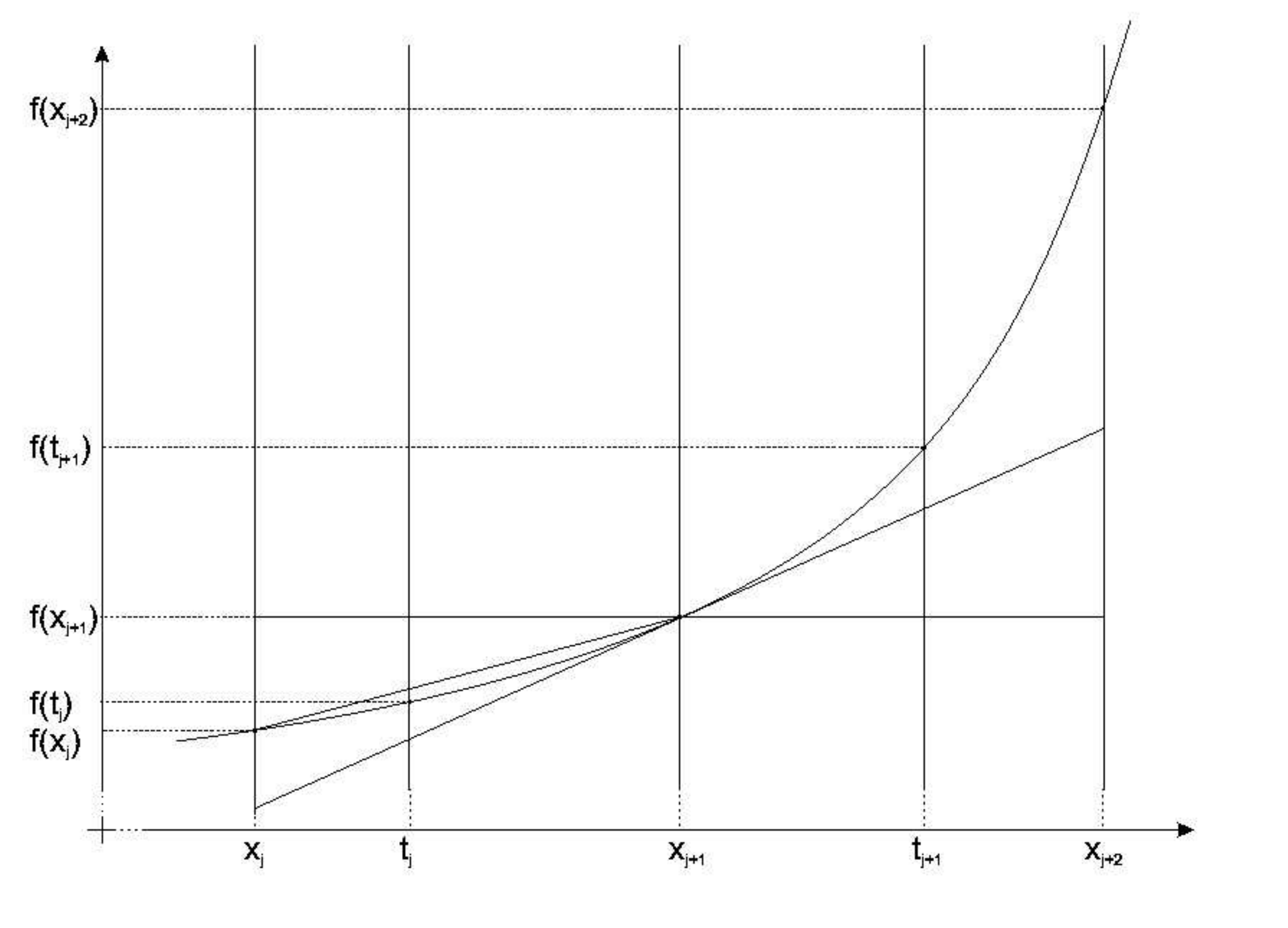}

From (\ref{med}) and (\ref{k1}), we have 
\begin{equation}
\Sigma _{1}=\frac{k_{0}}{2}-{\sum_{j=0}^{K-1}\left( k_{2j+1}-\frac{%
k_{2j}+k_{2j+2}}{2}\right) }+\frac{k_{2K}}{2}\geq \frac{k_{0}}{2}+\frac{%
k_{2K}}{2}  \label{lower}
\end{equation}%
and from (\ref{med}) and (\ref{k2}), we have 
\begin{equation}
\Sigma _{1}=k_{0}-\frac{k_{1}}{2}+{\sum_{j=1}^{K-1}\left( k_{2j}-\frac{%
k_{2j-1}+k_{2j+1}}{2}\right) }-\frac{k_{2K-1}}{2}+k_{2K}\leq k_{0}-\frac{%
k_{1}}{2}-\frac{k_{2K-1}}{2}+k_{2K}~.  \label{upper}
\end{equation}

Since equations (\ref{ss}) and (\ref{k2}) are related by definition of $%
\left\{ k_{j}\right\} _{j=0}^{2K}$ as%
\begin{equation*}
\Sigma _{1}=k_{0}-\frac{k_{1}}{2}+\Sigma -\frac{k_{2K-1}}{2}+k_{2K}
\end{equation*}%
the lower (\ref{lower}) and the upper (\ref{upper}) bounds yields 
\begin{equation*}
\frac{k_{0}}{2}+\frac{k_{2K}}{2}\leq k_{0}-\frac{k_{1}}{2}+\Sigma -\frac{%
k_{2K-1}}{2}+k_{2K}\leq k_{0}-\frac{k_{1}}{2}-\frac{k_{2K-1}}{2}+k_{2K}
\end{equation*}%
or, equivalently,%
\begin{equation*}
-\frac{k_{0}}{2}+\frac{k_{1}}{2}+\frac{k_{2K-1}}{2}-\frac{k_{2K}}{2}\leq
\Sigma \leq 0
\end{equation*}%
which, in view of definitions (\ref{kimpar}) and (\ref{kpar}), concludes the
proof of lemma.

\hfill $\Box $

\begin{corollary}
\label{cBorn} Let $f:\left[ a,b\right] \longrightarrow \mathbb{R}$ be a
concave function: 
\begin{equation*}
f\left( \lambda x+(1-\lambda )y\right) \geq \lambda f\left( x\right)
+(1-\lambda )f\left( y\right) \,
\end{equation*}%
for all $x,y\in \left[ a,b\right] $ and $0<\lambda <1$ and let $P$, $\left(
t_{j}\right) _{j=1}^{K}$ and $\Sigma (f;P)$ be as in the previous lemma.
Then 
\begin{equation*}
0\leq \Sigma (f;P)\leq \left( -\frac{f\left( t_{0}\right) }{2}+\frac{f\left(
x_{1}\right) }{2}+\frac{f\left( x_{K}\right) }{2}-\frac{f\left( t_{K}\right) 
}{2}\right) \Delta x.
\end{equation*}
\end{corollary}

\hfill $\Box $

\begin{remark}
The ideas of this proof was based in an argument used to establish the
phenomenon of Fresnel diffraction (see e.g. \cite{Born}).
\end{remark}

We are now in position to prove Theorem \ref{akna}.

\section{Proof of Theorem \protect\ref{akna} \label{PTakna}}

\setcounter{equation}{0} \setcounter{theorem}{0}

We shall proceed the asymptotic analysis applying the steepest descents
method to the integral kernel (\ref{KNzw}). For this we assume $N$ to be
large in comparison to all other variables which, from now on, are kept
fixed.

It is convenient rewrite $z$, $w$ and the difference of their argument using
scale parameters $\gamma $ and $\beta $: 
\begin{eqnarray}
Z &=&zN^{\gamma }  \notag \\
W &=&wN^{\gamma },\ \ \gamma >0  \label{escala}
\end{eqnarray}%
and 
\begin{equation}
\theta =N^{\beta }\left( \arg z-\arg w\right) ,\ \ \beta >0  \label{escala1}
\end{equation}%
The equation (\ref{KN}) can thus be written as 
\begin{equation}
K_{N}^{\alpha }\left( \frac{Z}{N^{\gamma }},\frac{W}{N^{\gamma }}\right) =%
\frac{\alpha }{2\pi }e^{-N^{1-\alpha \gamma }\left\vert Z\right\vert
^{\alpha }/2}e^{-N^{1-\alpha \gamma }\left\vert W\right\vert ^{\alpha }/2}%
\frac{N^{2\gamma }}{Z\overline{W}}S_{N}  \label{KaN}
\end{equation}%
where 
\begin{equation}
S_{N}:=\sum_{j=1}^{N}\frac{\left( N^{2(1-\alpha \gamma )/\alpha }Z\bar{W}%
\right) ^{j}}{\Gamma \left( 2j/\alpha \right) }.  \label{soma}
\end{equation}

We introduce another auxiliary scale parameter $\delta $ satisfying $%
0<\delta <1$ and 
\begin{equation}
\alpha \gamma +\delta =1~,  \label{alfagama}
\end{equation}%
in order to adjust the spacing in the label that indexes the sum. Note that $%
\gamma $ and $\delta $ are not independent. Equation (\ref{soma}) can be
written as 
\begin{equation}
S_{N}=\sum_{j=0}^{N-1}\frac{\left( N^{2\delta /\alpha }Z\bar{W}\right)
^{y_{j}N^{\delta }}}{\Gamma \left( 2y_{j}N^{\delta }/\alpha \right) }
\label{soma1}
\end{equation}%
where 
\begin{equation}
y_{j}=N^{-\delta }+jN^{-\delta }\ ,\qquad j=0,\ldots ,N-1~.  \label{tj}
\end{equation}

Given a function $f$ of the class $C^{\left( p\right) }$ in $\left[ a,b%
\right] $, the Euler-Maclaurin sum formula (see e.g. \cite{Abramovitz} with $%
\omega =0$ and $p=1$) 
\begin{equation}
\sum_{j=0}^{N-1}f\left( y_{j}\right) =\frac{1}{h}\int_{a}^{b}f\left(
x\right) dx+R_{1}+R_{2}  \label{emsf}
\end{equation}%
associated with the uniform partition $P:a=y_{0}<y_{1}<\cdots <y_{N}=b$,%
\begin{equation*}
y_{j}=a+jh,
\end{equation*}%
for $\,j\in \left\{ 0,\cdots ,N-1\right\} $, can be employed to estimate the
errors%
\begin{equation*}
R_{1}=\frac{1}{2}\left( f\left( b\right) -f\left( a\right) \right)
\end{equation*}%
and 
\begin{equation*}
R_{2}=-h\int_{0}^{1}\left( \frac{1}{2}-t\right) \left(
\sum_{j=0}^{N-1}f^{\prime }\left( a+\left( j+t\right) h\right) \right) dt
\end{equation*}%
in replacing the Darboux--Riemann sum of $f$ by its integral.

We take%
\begin{equation}
f(y)=g_{Z\overline{W}}\left( yN^{\delta }\right) =\frac{\left( N^{2\delta
/\alpha }Z\bar{W}\right) ^{yN^{\delta }}}{\Gamma \left( 2yN^{\delta }/\alpha
\right) },  \label{efe}
\end{equation}%
in (\ref{emsf}) with $g_{\zeta }(x)$ defined by (\ref{ge}). The partition $%
N^{-\delta }=y_{0}<y_{1}<\cdots <y_{N-1}=N^{1-\delta }$ of $\left[
N^{-\delta },N^{1-\delta }\right] $ is chosen with the $y_{j}$'s given by (%
\ref{tj}). In order to simplify the notation in (\ref{efe}), from now on we
fix $\zeta =Z\bar{W}=\left\vert \zeta \right\vert e^{i\theta /N^{\beta }}$.

Equation (\ref{soma}) can thus be written as 
\begin{equation}
S_{N}=N^{\delta }\int_{N^{-\delta }}^{N^{1-\delta }}g_{\zeta }\left(
N^{\delta }y\right) dy+r_{1}+r_{2},  \label{emsf2}
\end{equation}%
where 
\begin{equation}
r_{1}=\frac{1}{2}\left( g_{\zeta }\left( N\right) -g_{\zeta }\left( 1\right)
\right)  \label{r11}
\end{equation}%
and 
\begin{eqnarray}
r_{2} &=&-N^{-\delta }\int_{0}^{1}\left( \frac{1}{2}-t\right) \left(
\sum_{j=0}^{N-1}f^{\prime }\left( N^{-\delta }+\left( j+t\right) N^{-\delta
}\right) \right) dt  \notag \\
&=&-\sum_{j=0}^{N-1}\int_{0}^{1}\left( \frac{1}{2}-t\right) df\left( \left(
j+t\right) N^{-\delta }\right)  \notag \\
&=&-\sum_{j=0}^{N-1}\int_{0}^{1}\left( \frac{1}{2}-t\right) dg_{\zeta
}\left( j+t\right) .  \label{r22}
\end{eqnarray}

The proof now proceeds in two parts. The longest one, Part $\mathbf{I}$,
concerns with the estimates of $r_{1}$ and $r_{2}$. Part $\mathbf{II}$
applies the method of steepest descents to the integral term of the
representation (\ref{emsf}).

\medskip

\noindent $\mathbf{I}.$\textbf{\ Estimate of} $r_{1}$ \textbf{and} $r_{2}$%
\textbf{.} By the Stirling formula (see (\ref{gmx})), 
\begin{equation*}
g_{\zeta }\left( N\right) =\frac{\left( N^{2\delta /\alpha }\zeta \right)
^{N}}{\Gamma \left( 2N/\alpha \right) }=\sqrt{\frac{N}{\alpha \pi }}\left( 
\frac{\alpha e}{2N}\right) ^{2N/\alpha }\left( N^{\frac{2\delta }{\alpha }%
}\zeta \right) ^{N}(1+O\left( 1/N\right) )=O\left( N^{-k}\right)
\end{equation*}%
holds for any power $k$ of $1/N$, in view of $2N\left( 1-\delta \right)
/\alpha >0$. Since 
\begin{equation*}
g_{\zeta }\left( 1\right) =\frac{N^{2\delta /\alpha }\zeta }{\Gamma \left(
2/\alpha \right) }~
\end{equation*}%
we conclude, by (\ref{r11}), 
\begin{equation}
r_{1}=O\left( N^{2\delta /\alpha }\right) .  \label{r1}
\end{equation}

According to the second mean value theorem (see e.g. \cite{Rudin2}), for
each $j\in \left\{ 0,\ldots ,N-1\right\} $ there exist $x_{j}\in \left[ 0,1%
\right] $ such that 
\begin{equation*}
\int_{0}^{1}\left( \frac{1}{2}-t\right) dg_{\zeta }\left( j+t\right) =-\frac{%
1}{2}\left( g_{\zeta }\left( j+x_{j}\right) -g_{\zeta }\left( j\right)
\right) +\frac{1}{2}\left( g_{\zeta }\left( j+1\right) -g_{\zeta }\left(
j+x_{j}\right) \right)
\end{equation*}%
Taking this into consideration, (\ref{r22}) can thus be written as 
\begin{equation}
r_{2}=\frac{1}{2}\sum_{j=1}^{N}\left( 2g_{\zeta }\left( j+x_{j}\right)
-\left( g_{\zeta }\left( j\right) +g_{\zeta }\left( j+1\right) \right)
\right) ~.  \label{r2}
\end{equation}

Some considerations about (\ref{r2}) are required. We have to avoid to take
absolute value inside the sum since any estimate that disregards the change
of sign in (\ref{r2}), leads $r_{2}$ to be of the leading order of the
integral (\ref{emsf}) given by $O\left( N^{\delta }e^{N^{\delta }\left\vert
\zeta \right\vert ^{\alpha /2}}\right) $\footnote{%
We have $g_{\left\vert \zeta \right\vert }(x^{\ast })=O\left( N^{\delta
/2}e^{N^{\delta }\left\vert \zeta \right\vert ^{\alpha /2}}\right) $, the
Euler--Maclaurin formula (\ref{emsf2}) gives an extra $N^{\delta }$ and $%
N^{-\delta /2}$ results from the Gaussian integration in the
steepest--descent method. See Part $\mathbf{II.}$ for more detail.}. This
follows by (\ref{gm}) and the fact that there are $O(N^{\delta /2})$ terms
contributing to the sum (\ref{r2}), in view of Lemma \ref{distmax}. One
needs to be careful and exploit the change of sign in a clever way in order
to reduce the dependence on $N$ from the number of terms of this sum.
Because the estimates involve exponential growth, it is convenient to divide 
$r_{2}$ by the maximum value of $N^{\delta /2}g_{\left\vert \zeta
\right\vert }(x)$ (see (\ref{gm})). We set%
\begin{equation}
\hat{r}_{i}=\frac{r_{i}}{N^{\delta /2}g_{\left\vert \zeta \right\vert
}(x^{\ast })}  \label{rhat}
\end{equation}%
for $i=1,2$, and note by (\ref{r1}) that $\hat{r}_{1}$ is exponentially
small in $N^{\delta }$.

Writing $\zeta =\left\vert \zeta \right\vert e^{i\theta /N^{\beta }}$ with $%
\theta \in \mathbb{R}$, we have by definition (\ref{ge}) 
\begin{equation}
g_{\zeta }\left( x\right) =g_{\left\vert \zeta \right\vert }\left( x\right)
\cos \theta N^{-\beta }x+ig_{\left\vert \zeta \right\vert }\left( x\right)
\sin \theta N^{-\beta }x.  \label{RIm}
\end{equation}%
As $r_{2}$ is a linear function of $g_{\zeta }$, it suffices to estimate its
real part $\Re \mathrm{e}\left( r_{2}\right) $, since the estimate of $\Im 
\mathrm{m}(r_{2})$ can be done in analogous manner.

The estimation of the real and imaginary parts of (\ref{RIm}) depends on the
period 
\begin{equation}
p=\frac{2\pi }{\left\vert \theta \right\vert }N^{\beta }  \label{per}
\end{equation}%
of oscillation of $g_{\zeta }\left( x\right) $. For this, let $n_{N}(\theta
) $ be the cardinality of the set 
\begin{equation}
A_{N}(\theta )=\left\{ l\in \mathbb{N}:\frac{\left\vert \theta \right\vert }{%
\pi }N^{-\beta }<l\leq \frac{\left\vert \theta \right\vert }{\pi }N^{1-\beta
}\right\} .  \label{raiz2}
\end{equation}%
The number $n_{N}(\theta )$ counts how many oscillations between the maximum
and minimum value of $\cos \theta N^{-\beta }x$ there are as $x$ varies in
the interval $\left[ 1,N\right] $. For pedagogical reason, we divide the
estimate in two cases $\mathbf{(i)}$ $n_{N}(\theta )=O(1)$ and $\mathbf{(ii)}
$ $n_{N}(\theta )=O(N^{\varepsilon })$ for some $0<\varepsilon \leq 1-\beta $
\footnote{%
We set $\varepsilon =0$ when $\beta \geq 1$. In this case $n_{N}(\theta )$
is always $O(1)$. If $\beta <1$, $n_{N}(\theta )=O(1)$ when $\theta =O\left(
N^{-1+\beta }\right) $.}. The estimate for the first case can be done with
less effort. In the second case, which may also include the previous one,
the estimate is more subtle and leads to sharper result.

\medskip

\noindent $\mathbf{(i)}$ If $n_{N}(\theta )=n=O(1)$, we write (\ref{r2}) as 
\begin{equation*}
r_{2}=r_{2}^{(1)}+r_{2}^{(2)}
\end{equation*}%
where the real part of $r_{2}^{(i)}$, with $i=1,$ $2$, is given by 
\begin{equation}
\Re \mathrm{e}r_{2}^{(i)}=\sum_{j\in A_{N}^{(i)}}\left( \Re \mathrm{e}%
g_{\zeta }\left( j+x_{j}\right) -\left( \frac{\Re \mathrm{e}g_{\zeta }\left(
j\right) +\Re \mathrm{e}g_{\zeta }\left( j+1\right) }{2}\right) \right)
\label{rc}
\end{equation}%
with $A_{N}^{(i)}$ being the set of points $j\in \left\{ 1,\ldots ,N\right\} 
$ such that 
\begin{equation*}
\Re \mathrm{e}g_{\zeta }\left( j+1\right) -\Re \mathrm{e}g_{\zeta }\left(
j+x_{j}\right) \left\{ 
\begin{array}{cc}
\geq 0 & \mathrm{if}\ i=1 \\ 
<0 & \mathrm{if}\ i=2%
\end{array}%
\right. \,~.
\end{equation*}

Let $\left( j_{k}\right) _{k=1}^{L}$ denote a sequence of points right
before $\Re \mathrm{e}g_{\zeta }\left( j+1\right) -\Re \mathrm{e}g_{\zeta
}\left( j+x_{j}\right) $, as a function of $j\in \left\{ 1,\ldots ,N\right\} 
$, changes its sign:%
\begin{eqnarray*}
A_{N}^{(1)} &=&\left\{ 1,\ldots ,j_{1}\right\} \cup \left\{ j_{2}+1,\ldots
,j_{3}\right\} \cup \cdots \cup \left\{ j_{L-1}+1,\ldots ,j_{L}\right\} \\
A_{N}^{(2)} &=&\left\{ j_{1}+1,\ldots ,j_{2}\right\} \cup \left\{
j_{3}+1,\ldots ,j_{4}\right\} \cup \cdots \cup \left\{ j_{L}+1,\ldots
,N\right\} ~.
\end{eqnarray*}%
Since $0\leq x_{j}\leq 1$ and $g_{\left\vert \zeta \right\vert }\left(
x\right) $ is increasing in $\left[ 1,x^{\ast }\right) $ and decreasing in $%
\left( x^{\ast },N\right] $, the points $\left( j_{k}\right) _{k=1}^{L}$ are
essentially determined by the oscillations of the function $\cos \theta
N^{-\beta }x$ in $\Re \mathrm{e}g_{\zeta }\left( x\right) =g_{\left\vert
\zeta \right\vert }\left( x\right) \cos \theta N^{-\beta }x$ and $L=O\left(
n_{N}(\theta )\right) =O(1)$, by hypothesis.

By definition, we have%
\begin{eqnarray*}
\left\vert \Re \mathrm{e}r_{2}^{(1)}\right\vert &\leq &\frac{1}{2}\left\vert
\sum_{j\in A_{N}^{(1)}}\left( \Re \mathrm{e}g_{\zeta }\left( j+1\right) -\Re 
\mathrm{e}g_{\zeta }\left( j\right) \right) \right\vert \\
&=&\frac{1}{2}\left\vert \Re \mathrm{e}g_{\zeta }\left( j_{1}+1\right) -\Re 
\mathrm{e}g_{\zeta }\left( 1\right) +\cdots +\Re \mathrm{e}g_{\zeta }\left(
j_{L}+1\right) -\Re \mathrm{e}g_{\zeta }\left( j_{L-1}+1\right) \right\vert
\end{eqnarray*}%
and%
\begin{eqnarray*}
\left\vert \Re \mathrm{e}r_{2}^{(2)}\right\vert &<&\frac{1}{2}\left\vert
\sum_{j\in A_{N}^{(2)}}\left( \Re \mathrm{e}\left( g_{\zeta }\left( j\right)
\right) -\Re \mathrm{e}\left( g_{\zeta }\left( j+1\right) \right) \right)
\right\vert \\
&=&\frac{1}{2}\left\vert \Re \mathrm{e}g_{\zeta }\left( j_{1}+1\right) -\Re 
\mathrm{e}g_{\zeta }\left( j_{2}+1\right) +\cdots +\Re \mathrm{e}g_{\zeta
}\left( j_{L}+1\right) -\Re \mathrm{e}g_{\zeta }\left( N+1\right) \right\vert
\end{eqnarray*}%
so that%
\begin{equation*}
\left\vert \Re \mathrm{e}r_{2}\right\vert \leq \sum_{k=1}^{L}g_{\left\vert
\zeta \right\vert }\left( j_{k}+1\right) +\frac{g_{\left\vert \zeta
\right\vert }\left( 1\right) +g_{\left\vert \zeta \right\vert }\left(
N+1\right) }{2}~
\end{equation*}%
yields, together with (\ref{rhat}), (\ref{r1}), Lemma \ref{bor} and the fact
that the same holds for $\Im \mathrm{m}(r_{2})$, 
\begin{equation*}
\left\vert \hat{r}_{2}\right\vert \leq O\left( \frac{1}{N^{\delta /2}}%
\right) ~.
\end{equation*}

\medskip

\noindent $\mathbf{(ii)}$ Let $n_{N}(\theta )=O(N^{\varepsilon })$ for some $%
0<\varepsilon \leq 1-\beta $. Integrating (\ref{r22}) by parts gives 
\begin{eqnarray}
r_{2} &=&\sum_{j=1}^{N}\int_{0}^{1}\left( \frac{1}{2}-t\right) dg_{\zeta
}(j+t)  \notag \\
&=&\sum_{j=1}^{N}\left( \left. \left( \frac{1}{2}-t\right) g_{\zeta
}(j+t)\right\vert _{0}^{1}+\int_{0}^{1}g_{\zeta }(j+t)dt\right)  \notag \\
&=&\sum_{j=1}^{N}\left( \int_{0}^{1}g_{\zeta }(j+t)dt-\frac{1}{2}\left(
g_{\zeta }(j)+g_{\zeta }(j+1)\right) \right) .  \label{r223}
\end{eqnarray}%
We now split the above sum into 
\begin{equation}
r_{2}=r_{2}^{\sqcup }+r_{2}^{\sqcap }  \label{sss}
\end{equation}%
where the real part of $r_{2}^{\sqcup (\sqcap )}$ is given by 
\begin{equation*}
\Re \mathrm{e}r_{2}^{\sqcup (\sqcap )}=\sum_{j\in A_{N}^{\sqcup \left(
\sqcap \right) }}\left( \int_{0}^{1}\Re \mathrm{e}g_{\zeta }\left(
j+t\right) dt-\frac{1}{2}\left( \Re \mathrm{e}g_{\zeta }\left( j\right) +\Re 
\mathrm{e}g_{\zeta }\left( j+1\right) \right) \right)
\end{equation*}%
with $A_{N}^{\sqcup \left( \sqcap \right) }$ being the set of points $j\in
\left\{ 1,\cdots ,N\right\} $ such that $\left( \Re \mathrm{e}g_{\zeta
}\right) ^{\prime \prime }\left( j\right) \geq 0\ (<0)$.

Let us note that the function $\Re \mathrm{e}g_{\zeta }(x)=$ $g_{\left\vert
\zeta \right\vert }\left( x\right) \cos \theta x$ always has a well defined
concavity and the cardinality of inflection points is of same order in $N$
of the cardinality of critical points, since the main function responsible
for both, the number of oscillations and changes of concavity, is the cosine.

Let $\left( k_{i}\right) _{i=1}^{L}$ denote a sequence of points in $\left\{
1,\ldots ,N\right\} $ right before $\left( \Re \mathrm{e}g_{\zeta }\right)
^{\prime \prime }\left( j\right) $ changes sign. Analogously, we have 
\begin{eqnarray*}
A_{N}^{\sqcap } &=&\left\{ 1,\ldots ,k_{1}\right\} \cup \left\{
k_{2}+1,\ldots ,k_{3}\right\} \cup \cdots \cup \left\{ k_{L-1}+1,\ldots
,k_{L}\right\} \\
A_{N}^{\sqcup } &=&\left\{ k_{1}+1,\ldots ,k_{2}\right\} \cup \left\{
k_{3}+1,\ldots ,k_{4}\right\} \cup \cdots \cup \left\{ k_{L}+1,\ldots
,N\right\} ~
\end{eqnarray*}%
where, by the same reason as in item $\mathbf{(i)}$, $L=O\left( n_{N}(\theta
)\right) =O(N^{\varepsilon })$ and, consequently, 
\begin{equation}
k_{i+1}-k_{i}=O\left( N^{1-\varepsilon }\right)  \label{jj}
\end{equation}%
holds for $i=1,\ldots ,L-1$. Note also that, by (\ref{per}), 
\begin{equation}
\theta =O\left( N^{\varepsilon +\beta -1}\right)  \label{Oteta}
\end{equation}

Applying Lemma \ref{Born} (and Corollary \ref{cBorn}) to each interval $%
I_{i}=\left\{ k_{i}+1,\ldots ,k_{i+1}\right\} $, $i=0,\ldots ,L$ ($%
k_{0}\equiv 0$ and $k_{L+1}=N$) of size $K=O\left( N^{1-\varepsilon }\right) 
$ with $f(x)$ replaced by $\Re \mathrm{e}g_{\zeta }(x)$ and $\Delta =1$,
yields%
\begin{equation}
\left\vert \Re \mathrm{e}r_{2}\right\vert \leq \sum_{i=1}^{L}\left\vert \Re 
\mathrm{e}g_{\zeta }\left( t_{i}\right) -\frac{\Re \mathrm{e}g_{\zeta
}(k_{i})+\Re \mathrm{e}g_{\zeta }\left( k_{i}+1\right) }{2}\right\vert
+\left\vert \Re \mathrm{e}g_{\zeta }\left( 1\right) \right\vert +\left\vert
\Re \mathrm{e}g_{\zeta }\left( N+1\right) \right\vert ~,  \label{Rer2}
\end{equation}%
with $t_{k}$ defined by the mean value theorem $\Re \mathrm{e}g_{\zeta
}\left( t_{i}\right) =\displaystyle\int_{k_{i}}^{k_{i}+1}\Re \mathrm{e}%
g_{\zeta }(x)dx$. Note that the points $\left( k_{i}\right) _{i=1}^{L}$ are
closed to the inflection points $\left( x_{i}\right) _{i=1}^{L}$ of $\Re 
\mathrm{e}g_{\zeta }(x)$ and, moreover, the value of $\Re \mathrm{e}g_{\zeta
}(x)$ at these points are small compared with the maximum value $g_{|\zeta
|}(x^{\ast })$. We shall estimate the order of $\Re \mathrm{e}g_{\zeta
}(x_{i})$ and use Lemma \ref{distmax} to reduce the number of terms involved
in the sum (\ref{sss}).

Taking the second derivative of the real part of (\ref{RIm}), we obtain 
\begin{equation*}
\left( \Re \mathrm{e}g\right) ^{\prime \prime }\left( x\right) =\left(
g_{\left\vert \zeta \right\vert }^{\prime \prime }\left( x\right) -\theta
^{2}N^{-2\beta }g_{\left\vert \zeta \right\vert }\left( x\right) \right)
\cos \theta N^{-\beta }x-2\theta N^{-\beta }g_{\left\vert \zeta \right\vert
}^{\prime }\left( x\right) \mathrm{\sin }\theta N^{-\beta }x
\end{equation*}%
Since derivatives of $g_{\left\vert \zeta \right\vert }\left( x\right) $
increases its value by a logarithm of $N$ factor (see equation (\ref{glinha}%
)), combined with (\ref{Oteta}), it gives 
\begin{equation}
\frac{\left( \Re \mathrm{e}g\right) ^{\prime \prime }\left( x\right) }{%
g_{\left\vert \zeta \right\vert }\left( x\right) }=\left( O(\log
^{2}N)+O\left( N^{2(\varepsilon -1)}\right) \right) \cos \theta N^{-\beta
}x+O\left( N^{\varepsilon -1}\log N\right) \sin \theta N^{-\beta }x~.
\label{pi}
\end{equation}%
But we have, on the other hand,%
\begin{equation*}
\left( \Re \mathrm{e}g\right) ^{\prime \prime }\left( x_{i}\right) =\left(
g_{\left\vert \zeta \right\vert }^{\prime \prime }\left( x_{i}\right)
-\theta ^{2}N^{-2\beta }g_{\left\vert \zeta \right\vert }\left( x_{i}\right)
\right) \cos \theta N^{-\beta }x_{i}-2\theta N^{-\beta }g_{\left\vert \zeta
\right\vert }^{\prime }\left( x_{i}\right) \sin \theta N^{-\beta }x_{i}=0
\end{equation*}%
holds at each inflection point $x_{i}$. This together with (\ref{pi})
implies that the inflection point $x_{i}$ must be at $O(1/\log N)$ distance
from the $k$--th zero of $\cos \theta N^{-\beta }x$. Indeed, defining $%
\Delta _{i}=O\left( 1/\log N\right) $ by 
\begin{equation*}
x_{i}=\frac{(2i-1)\pi }{2\left\vert \theta \right\vert N^{-\beta }}+\Delta
_{i}
\end{equation*}%
we have 
\begin{eqnarray*}
\cos \theta N^{-\beta }x_{i} &=&\cos \left( \pm (i-1/2)\pi +\theta N^{-\beta
}\Delta _{k}\right) =\pm (-1)^{i}\sin \theta N^{-\beta }\Delta _{i}=O\left(
N^{\varepsilon -1}/\log N\right) \\
\sin \theta N^{-\beta }x_{i} &=&\sin \left( \pm (i-1/2)\pi +\theta N^{-\beta
}\Delta _{i}\right) =\mp (-1)^{i}\cos \theta N^{-\beta }\Delta _{i}=O\left(
1\right)
\end{eqnarray*}%
and, together with (\ref{pi}), one sees that $\left( \Re \mathrm{e}g\right)
^{\prime \prime }\left( x_{i}\right) =0$ holds in the leading order. Since
the points $k_{i}$, $t_{i}$ and $k_{i}+1$ are not distant from the
inflection point $x_{i}$ ($g_{\zeta }\left( x\right) $ varies slowly for
each interval $k_{i}\leq x\leq k_{i}+1$), 
\begin{equation}
\frac{\left\vert \Re \mathrm{e}g_{\zeta }\left( x\right) \right\vert }{%
g_{|\zeta |}\left( x\right) }=\left\vert \cos \theta N^{-\beta }x\right\vert
\leq O\left( N^{-1+\varepsilon }/\log N\right)  \label{gg}
\end{equation}%
holds for $x$ at the values $\left\{ k_{i},t_{i},k_{i}+1\right\} _{i=1}^{L}$.

The number of terms that contributes to (\ref{sss}), as well as to the sum (%
\ref{Rer2}), can be estimated using Lemma \ref{distmax}. Instead of an
interval $I$ of size $N$ we shall consider an interval $I^{\prime }$
containing $x^{\ast }$ with $O(N^{\delta /2}\log N)$ points. By (\ref{jj}),
a number of order $\left. N^{\delta /2}\log N\right/ N^{1-\varepsilon }$ of
terms give appreciably contribution to (\ref{Rer2}) and, together with (\ref%
{gg}), the fact that the same estimate holds for $\Im \mathrm{m}g_{\zeta
}\left( x\right) $ and (\ref{rhat}), we conclude 
\begin{equation*}
\left\vert \hat{r}_{2}\right\vert =O\left( N^{-2(1-\varepsilon )}\right)
\end{equation*}%
uniformly in every closed interval of $0<\varepsilon \leq 1-\beta $.

\medskip

\noindent $\mathbf{II}$\textbf{. The Method of Steepest Descents } Equation (%
\ref{emsf2}) can be written as 
\begin{equation}
S_{N}=N^{\delta /2}g_{\left\vert \zeta \right\vert }(x^{\ast })\left(
N^{\delta /2}\int_{N^{-\delta }}^{N^{1-\delta }}f(y)dy+\hat{r}_{1}+\hat{r}%
_{2}\right)  \label{S}
\end{equation}%
where, by the Stirling formula (see (\ref{gmx})),%
\begin{equation}
f\left( y\right) =\frac{g_{\zeta }\left( N^{\delta }y\right) }{g_{\left\vert
\zeta \right\vert }(x^{\ast })}=\sqrt{\frac{2y}{\alpha \left\vert \zeta
\right\vert ^{\alpha /2}}}e^{N^{\delta }h(y)}(1+O\left( 1/N^{\delta }\right)
)  \label{fify}
\end{equation}%
with%
\begin{equation}
h(y)=\frac{2y}{\alpha }\log \frac{\alpha e\zeta ^{\alpha /2}}{2y}-\left\vert
\zeta \right\vert ^{\alpha /2}  \label{h}
\end{equation}%
Note that $\Re \mathrm{e}h(y)\leq 0$ holds for all $y>0$ and attains to its
maximum $\Re \mathrm{e}h(y^{\ast })=0$ at $y^{\ast }=\alpha \left\vert \zeta
\right\vert ^{\alpha /2}/2$ inside the domain of integration $\left[
N^{-\delta },N^{1-\delta }\right] $, by condition $0<\delta <1$ and $N$
large enough.

We now use the steepest descents technique to estimate the integral that
appears in (\ref{S}). This technique uses the Cauchy theorem to deform the
interval of integration $\left[ N^{-\delta },N^{1-\delta }\right] $ into a
curve $\mathcal{C}$: 
\begin{equation}
I=\sqrt{\frac{2N^{\delta }}{\alpha \left\vert \zeta \right\vert ^{\alpha /2}}%
}\int_{N^{-\delta }}^{N^{1-\delta }}\sqrt{y}e^{N^{\delta }h(y)}dy=\sqrt{%
\frac{2N^{\delta }}{\alpha \left\vert \zeta \right\vert ^{\alpha /2}}}\int_{%
\mathcal{C}}\sqrt{\eta }e^{N^{\delta }h\left( \eta \right) }d\eta
\label{Inps}
\end{equation}%
where $h:\mathbb{C}\longrightarrow \mathbb{C}$ is extended analytically to
the complex plane, $\eta =y+iw$ and $\mathcal{C}$ is a smooth curve with
extreme points $\eta _{1}=N^{-\delta }$ and $\eta _{2}=N^{1-\delta }$ chosen
in such a way that $(a)$ it passes by the saddle point $\eta _{0}=\alpha
\zeta ^{\alpha /2}/2$ ($\left\vert \eta _{0}\right\vert =y^{\ast }$) defined
implicitly by 
\begin{equation}
h^{\prime }\left( \eta _{0}\right) =\frac{2}{\alpha }\log \frac{\alpha \zeta
^{\alpha /2}}{2\eta _{0}}=0  \label{hl}
\end{equation}%
and $(b)$ it maximizes the function $\Re \mathrm{e}h\left( y,w\right) $
along a level curve 
\begin{equation*}
\Im \mathrm{m}h\left( y,w\right) =c
\end{equation*}%
in a neighborhood $U_{0}$ of $\eta _{0}$. If, in addition, 
\begin{equation}
\Re \mathrm{e}h\left( y,w\right) \geq \max \{\Re \mathrm{e}h\left(
N^{-\delta },0\right) ,\Re \mathrm{e}h\left( N^{1-\delta },0\right) \}
\label{hh}
\end{equation}%
holds along $\mathcal{C}$, then the main contribution to (\ref{Inps}) will
be given by the saddle point $\eta _{0}$; if, on the other hand, (\ref{hh})
cannot be satisfied to any such curve $\mathcal{C}$, the main contribution
to the integral (\ref{Inps}) will be given by the extreme points.

At the extreme points, neither $\eta _{1}$ nor $\eta _{2}$ plays an
important role, since both leave the integral (\ref{Inps}) exponentially
small with $N$. So, the contribution to (\ref{Inps}) is given by the
vicinity of the saddle point.

Expanding $h$ in Taylor series about $\eta _{0}=\alpha \zeta ^{\alpha
/2}/2=\alpha \left\vert \zeta \right\vert ^{\alpha /2}e^{i\alpha N^{-\beta
}\theta /2}/2$, gives 
\begin{eqnarray*}
h\left( \eta \right) &=&h\left( \eta _{0}\right) +\frac{1}{2}h^{\prime
\prime }\left( \eta _{0}\right) \left( \eta -\eta _{0}\right) ^{2}+O\left(
\left( \eta -\eta _{0}\right) ^{3}\right) \\
&=&\zeta ^{\alpha /2}-\left\vert \zeta \right\vert ^{\alpha /2}-\frac{2}{%
\alpha ^{2}\left\vert \zeta \right\vert ^{\alpha /2}}\rho ^{2}e^{i(2\varphi
-\alpha \theta N^{-\beta }/2)}+O\left( \left( \eta -\eta _{0}\right)
^{3}\right)
\end{eqnarray*}%
with $\eta -\eta _{0}=\rho e^{i\varphi }\in U_{0}$. We choose $\mathcal{C}$
so that $2\varphi -\alpha \theta N^{-\beta }/2=0$ at the the saddle point.
Applying the steepest descents technique, the integral (\ref{Inps}) can be
approximate by a Gaussian integral in the vicinity $U_{0}$ of $\eta _{0}$,
resulting (see e.g. \cite{Murray}, for details)%
\begin{eqnarray}
S_{N} &=&N^{\delta /2}g_{\left\vert \zeta \right\vert }(x^{\ast })\left(
e^{N^{\delta }\left( \zeta ^{\alpha /2}-\left\vert \zeta \right\vert
^{\alpha /2}\right) }\sqrt{\frac{2N^{\delta }}{\alpha \left\vert \zeta
\right\vert ^{\alpha /2}}}\sqrt{\frac{2\pi \eta _{0}}{-N^{\delta }h^{\prime
\prime }\left( \eta _{0}\right) }}\left( 1+O\left( \frac{1}{N^{\delta }}%
\right) \right) +\hat{r}_{1}+\hat{r}_{2}\right)  \notag \\
&=&N^{\delta /2}g_{\left\vert \zeta \right\vert }(x^{\ast })\left(
e^{N^{\delta }\left( \zeta ^{\alpha /2}-\left\vert \zeta \right\vert
^{\alpha /2}\right) }\sqrt{\frac{2\pi }{\left\vert \zeta \right\vert
^{\alpha /2}}}\eta _{0}\left( 1+O\left( \frac{1}{N^{\delta }}\right) \right)
+\hat{r}_{1}+\hat{r}_{2}\right)  \label{S1}
\end{eqnarray}

Now, since by (\ref{N>}) $\alpha N^{\delta }\left\vert \zeta \right\vert
^{\alpha /2}/2<N$, 
\begin{eqnarray}
\left\vert \exp \left( N^{\delta }(\zeta ^{\alpha /2}-\left\vert \zeta
\right\vert ^{\alpha /2})\right) \right\vert &=&\exp \left( N^{\delta
}\left\vert \zeta \right\vert ^{\alpha /2}(\cos \alpha N^{-\beta }\theta
/2-1)\right)  \notag \\
&\geq &\exp \left( -\alpha \theta ^{2}N^{1-2\beta }\right)  \label{theta}
\end{eqnarray}%
and, provided $\beta \geq 1/2$, it follows from the estimates of $r_{1}$ and 
$r_{2}$ in $\mathbf{I.}$ that\newline
\begin{equation*}
S_{N}=\frac{\alpha }{2}\zeta ^{\alpha /2}N^{\delta }\exp \left( \zeta ^{%
\frac{\alpha }{2}}N^{\delta }\right) \left( 1+E_{N}^{\alpha .\delta }(\zeta
)\right)
\end{equation*}%
with%
\begin{equation*}
\left\vert E_{N}^{\alpha .\delta }(\zeta )\right\vert \leq O\left(
N^{-\delta /2}\right)
\end{equation*}%
whenever $\zeta \in S(\theta N^{-1/2},K^{\alpha ,\delta })$, where $%
K^{\alpha ,\delta }=\left( 2N^{1-\delta }/\alpha \right) ^{2/\alpha }$.
Therefore, we obtain from (\ref{KaN}) 
\begin{equation}
\frac{1}{N^{\delta +2\gamma }}K_{N}^{\alpha }\left( \frac{Z}{N^{\gamma }},%
\frac{W}{N^{\gamma }}\right) =\frac{\alpha ^{2}}{4\pi }\left( Z\bar{W}%
\right) ^{\frac{\alpha }{2}-1}e^{N^{\delta }\left( \left( Z\bar{W}\right) ^{%
\frac{\alpha }{2}}-\frac{\left\vert Z\right\vert ^{\alpha }}{2}-\frac{%
\left\vert W\right\vert ^{\alpha }}{2}\right) }\left( 1+E_{N}^{\alpha
.\delta }(Z\bar{W})\right)  \label{fkg}
\end{equation}%
where we have used (\ref{alfagama}) with $0<\delta <1$. In particular,
taking $\delta \nearrow 1$, 
\begin{equation}
\frac{1}{N}K_{N}^{\alpha }\left( Z,W\right) =\frac{\alpha ^{2}}{4\pi }\left(
Z\bar{W}\right) ^{\frac{\alpha }{2}-1}e^{N\left( \left( Z\bar{W}\right) ^{%
\frac{\alpha }{2}}-\frac{\left\vert Z\right\vert ^{\alpha }}{2}-\frac{%
\left\vert W\right\vert ^{\alpha }}{2}\right) }\left( 1+E_{N}^{\alpha .1}(Z%
\bar{W})\right) .  \label{nucleo}
\end{equation}

\hfill $\Box $

\begin{remark}
\label{scale}Equation (\ref{theta}) prevents $\zeta =Z\bar{W}$ to be defined
in a sector $S(\theta N^{-\beta },K^{\alpha ,\delta })$ of opening wider
than $O(N^{-1/2})$. The introduction of the scale $\delta <1$ guarantees
that the main contribution to (\ref{Inps}) comes from the saddle point for
any $\zeta \in \mathbb{C}$ fixed. Note that $K^{\alpha ,\delta
}=O(N^{2(1-\delta )/\alpha })$ and for $\delta =1$ we need $\left\vert \zeta
\right\vert \leq K^{\alpha ,1}=(2/\alpha )^{2/\alpha }$ (see Remark \ref%
{bound}). As the calculation in the appendix below indicates, $\left\vert
\zeta \right\vert $ may be even smaller than that, depending on the sector
opening $\tau $.
\end{remark}

\appendix%

\section{Taylor Remainder\label{TR}}

\setcounter{equation}{0} \setcounter{theorem}{0}

Let $f_{N}(\zeta )=N\zeta e^{N\zeta }$ be a function defined for $\zeta
=\left\vert \zeta \right\vert e^{i\theta }\in \mathbb{C}$ and $N$ a fixed
natural number. Its Taylor remainder with respect to the polynomial $%
S_{N}(\zeta )=N\zeta +\cdots +\dfrac{1}{(N-1)!}(N\zeta )^{N}$ of order $N$
can be expressed by the Lagrange formula (see e.g. \cite{T}) 
\begin{equation*}
R_{N}(\zeta )=f_{N}(\zeta )-S_{N}(\zeta )=\frac{1}{(N+1)!}g_{N}^{(N+1)}(a)
\end{equation*}%
for some $0<a<1$, where $g_{N}(x)=f_{N}(x\zeta )$, $x\in \left[ 0,1\right] $%
, satisfies%
\begin{equation}
g_{N}^{(r)}(x)=\left( rN^{r}+N^{r+1}x\zeta \right) \zeta ^{r}e^{Nx\zeta }~
\label{gr}
\end{equation}%
for every $r\in \mathbb{N}$, by induction.

Writing 
\begin{equation*}
S_{N}(\zeta )=f_{N}(\zeta )(1+E_{N}(\zeta ))
\end{equation*}%
the error function $E_{N}(\zeta )=R_{N}(\zeta )/f_{N}(\zeta )$ is estimated
for $\zeta $ in a sectorial domain $S(\tau ,K)=\left\{ \zeta \in \mathbb{C}%
:\left\vert \arg (\zeta )\right\vert <\tau /2\,,\ \left\vert \zeta
\right\vert <K\right\} $ using (\ref{gr}) together with the Stirling formula 
$r!=\sqrt{2\pi r}(r/e)^{r}(1+O(1/r))$: 
\begin{equation*}
\left\vert E_{N}(\zeta )\right\vert =\frac{1}{\sqrt{2\pi N}}\left\vert
1+a\zeta \right\vert e^{N}\left\vert \zeta \right\vert
^{N}e^{-N(1-a)\left\vert \zeta \right\vert \cos \theta }(1+O(1/N))
\end{equation*}%
so $\sup_{\zeta \in S(\tau ,K)}\left\vert E_{N}(\zeta )\right\vert =O\left(
1/\sqrt{N}\right) $ where $K=K(a,\tau )>0$ is given by the smallest
solutions of 
\begin{equation}
Ke^{-(1-a)K\cos \tau /2+1}=1~,~  \label{eqK}
\end{equation}%
which exists and is continuous for all $0<a<1$ and $\tau \in \lbrack 0,2\pi
] $. The implicit solutions of (\ref{eqK}) for $K=K(a,\tau )$ are described
in figure below for $a=1/2$, $1/8$ and $1/16$.

\includegraphics{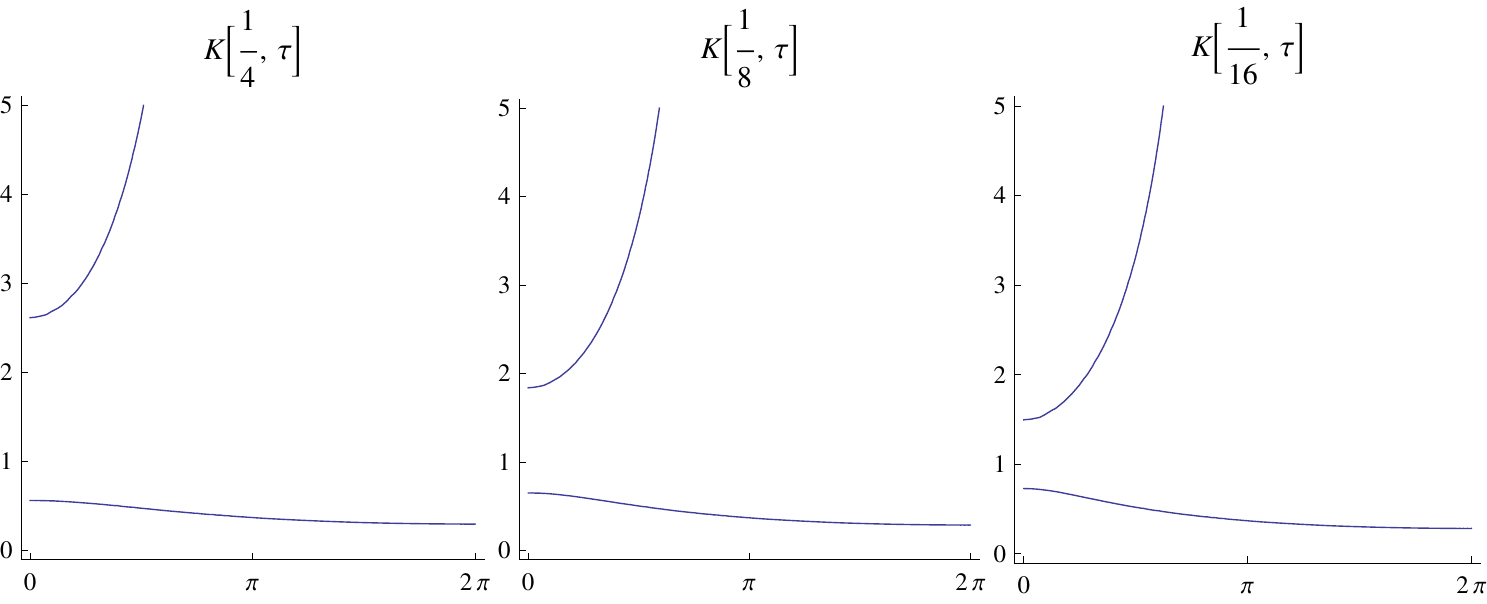}

\section{Proof of Corollary \protect\ref{npoint}\label{PCnpoint}}

\setcounter{equation}{0} \setcounter{theorem}{0}

Assuming temporarily that (\ref{kd1}) holds with $Z=W=r$, we observe that by
(\ref{Zi}) 
\begin{eqnarray*}
Z_{i} &=&r+\frac{1}{\sqrt{N}}\frac{2z_{i}}{\alpha \left\vert r\right\vert
^{\alpha /2-1}}+O(1/N) \\
&=&r\exp \left( \frac{1}{\sqrt{N}}\frac{2z_{i}}{\alpha \left\vert
r\right\vert ^{\alpha /2-1}}+O(1/N)\right)
\end{eqnarray*}%
and 
\begin{equation*}
\arg \left( Z_{i}\bar{Z}_{j}\right) <\theta /\sqrt{N}~,
\end{equation*}%
for some $\theta >0$ and any $i$, $j$, if $N$ is large enough, say $N>N_{1}$%
. We take, in addition, $N>N_{0}$ where $N_{0}$ is given by (\ref{N>}) with $%
1/\left\vert \zeta \right\vert $ and $\left\vert \zeta \right\vert $
replaced by $1/\min_{i,j}\left( \left\vert Z_{i}\bar{Z}_{j}\right\vert
\right) $ and $\max_{i,j}\left( \left\vert Z_{i}\bar{Z}_{j}\right\vert
\right) $, respectively. So, for $N>\max (N_{0},N_{1})$ equation (\ref{kd1})
holds with $(r,r)$ and $\left( Z_{i},Z_{j}\right) $, for any $i,j$, in the
place of $\left( Z,W\right) $. From equation (\ref{E}) and (\ref{Zi}), it
holds for $r\in \mathbb{C}$ with $0<\left\vert r\right\vert <(2/\alpha
)^{1/\alpha }$, whose closure is the support of the eigenvalues density (see
eq. \ref{ro}).

Now, applying the Taylor expansion%
\begin{equation*}
(1+w)^{\alpha /2}=1+\frac{\alpha }{2}w+\frac{\alpha }{4}\left( \frac{\alpha 
}{2}-1\right) w^{2}+O(w^{3})
\end{equation*}%
to the exponent of $K_{N}^{\alpha }\left( Z_{i},Z_{j}\right) $, yields 
\begin{equation}
N\left( \left( Z_{i}\bar{Z}_{j}\right) ^{\alpha /2}-\frac{1}{2}\left\vert
Z_{i}\right\vert ^{\alpha }-\frac{1}{2}\left\vert Z_{j}\right\vert ^{\alpha
}\right) =A_{ij}+i\sqrt{N}B_{ij}+O(1/\sqrt{N})  \label{AB}
\end{equation}%
where%
\begin{eqnarray*}
A_{ij} &=&z_{i}\bar{z}_{j}-\frac{1}{2}\left\vert z_{i}\right\vert ^{2}-\frac{%
1}{2}\left\vert z_{j}\right\vert ^{2}~, \\
B_{ij} &=&\lambda _{i}-\lambda _{j}
\end{eqnarray*}%
and%
\begin{equation*}
\lambda _{i}=\left\vert r\right\vert ^{\alpha /2+1}\Im \mathrm{m}\frac{z_{i}%
}{r}+\frac{1}{2\sqrt{N}}\left\vert r\right\vert ^{2}\left( 1-\frac{2}{\alpha 
}\right) \Im \mathrm{m}\frac{z_{i}^{2}}{r^{2}}~
\end{equation*}%
is a real number. Let $C_{N}$ and $D_{N}$ denote $n\times n$ matrices with
respective entries $\left( C_{N}\right) _{ij}=\dfrac{1}{\pi }\exp \left(
A_{ij}+i\sqrt{N}B_{ij}\right) (1+O(1/\sqrt{N}))$ and $D_{ij}=\dfrac{1}{\pi }%
\exp \left( A_{ij}\right) (1+O(1/\sqrt{N}))$ ($=C_{ij}$ with $B_{ij}=0$). If
we write $\Lambda _{N}=\text{diag}\left( \exp (i\sqrt{N}\lambda _{i})\right) 
$, then $C_{N}=\Lambda _{N}D_{N}\bar{\Lambda}_{N}$, $\Lambda _{N}\bar{\Lambda%
}_{N}=I$ ($\bar{\Lambda}_{N}$ and $I$ are the complex conjugate of $\Lambda
_{N}$ and the identity matrix) and 
\begin{equation*}
\det C_{N}=\det \Lambda _{N}D_{N}\bar{\Lambda}_{N}=\det D_{N}\bar{\Lambda}%
_{N}\Lambda _{N}=\det D_{N}~.
\end{equation*}%
by Cauchy-Binet formula. This concludes the proof since, by (\ref{Rn}) (\ref%
{kd1}) and (\ref{AB}), the l.h.s of (\ref{R}) is the determinant of a matrix
whose asymptotic expansion is given by $C_{N}$ and%
\begin{equation*}
\lim_{N\rightarrow \infty }\det C_{N}=\lim_{N\rightarrow \infty }\det
D_{N}=\det \left( \mathbb{K}\left( z_{i},z_{j}\right) \right) _{i,j=1}^{n}~
\end{equation*}%
by continuity.

\hfill $\Box $

\begin{center}
\textbf{Acknowledgements}
\end{center}

We would like to express our gratitude to Walter Wreszinski for his comments and advices. 
DHUM thanks Gordon Slade for his hospitality at UBC.


\begin{thebibliography}{99}
\bibitem{Abramovitz} M. Abramovitz, I. A. Stegun. \textquotedblleft Handbook
of Mathematical Functions\textquotedblright . Dover, New York 1970

\bibitem{Bergman} Stefan Bergman. \textquotedblleft The Kernel Function and
Conformal Mapping\textquotedblright , Mathematical Surveys and Monographs,
Vol. \textbf{5}, second ed., AMS 1970

\bibitem{Born} M. Born, E. Wolf. \textquotedblleft Principles of
optics\textquotedblright . Pergamon Press 1959

\bibitem{Chau Yu} L. L. Chau, Y. Yue. \textquotedblleft Unitary Polynomials
in normal matrix models and wave functions for the fractional quantum Hall
effects\textquotedblright . Phys. Lett. A \textbf{167}, 452 (1992)

\bibitem{Chau Zaboronsky} L. L. Chau, O. Zaboronsky. \textquotedblleft On
the Structure of Correlation Function in the normal matrix
models\textquotedblright . Comm. Math. Phys. \textbf{196 }, 203-247 (1998)

\bibitem{Deift} P. Deift. \textquotedblleft Orthogonal Polynomials and
Random Matrices: A Riemann-Hilbert Approach\textquotedblright . American
Mathematical Society. Courant Institute of Mathematical Sciences, New York
University, New York. Lecture Notes 3, AMS 2000

\bibitem{Donoghue} W. F. Donoghue Jr.. \textquotedblleft Monotone Matrix
Function and Analytic Continuation\textquotedblright , Die Grundlehren der
mathematischen Wissenschaften \textbf{207}, Springer-Verlag 1974

\bibitem{ElbauFelder} P. Elbau; G. Felder. "Density of Eigenvalues of Random
Normal Matrices", Comm. Math. Phys. \textbf{259}, 433-450 (2005)

\bibitem{Fyodorov1} Y. V. Fyodorov, H.-J. Sommers and B. A. Khoruzhenko.
\textquotedblleft Universality in the Random Matrix Spectra in the Regime of
Weak Non-Hermiticity \textquotedblright . Ann. Inst. Henri Poincar\'{e} 
\textbf{68}, n$^{0}$ 4 , 449-489 (1998)

\bibitem{Ginibre} J. Ginibre. \textquotedblleft Statistical Ensembles of
Complex, Quaternion and Real Matrices\textquotedblright . Journ. Math. Phys. 
\textbf{6}, Issue 3, pp. 440-449, (1965). Pac. Jour. Math. \textbf{193} , n$%
^{0}$ 2, 355-369 (2000)

\bibitem{Hedenmalm} H. Hedenmalm, N. Makarov. \textquotedblleft Quantum
Hele-Shaw flow\textquotedblright . arXiv: math. PR/0411437 v1 19/11/2004

\bibitem{Lancaster} P. Lancaster and M. Tismenetsky. Theory of Matrices, $%
2^{nd}$ edition with applications. Academic Press, San Diego (1985)

\bibitem{LL} Eli Levin, Doron S. Lubinsky. \textquotedblleft Universality
limits for exponential weights\textquotedblright . Constr. Approx. \textbf{29%
}, 247-275 (2009).

\bibitem{Murray} J. D. Murray. \textquotedblleft Asymptotic
analysis\textquotedblright . Clarendon Press - Oxford 1974

\bibitem{Rudin2} W. Rudin. \textquotedblleft Principles of mathematical
analysis\textquotedblright . McGraw-Hill 1964

\bibitem{Ruelle} D. Ruelle. \textquotedblleft Statistical Mechanics:
Rigorous Results\textquotedblright . Addison-Wesley Publishing Company 1989

\bibitem{Saff} E. B. Saff, V. Totik. \textquotedblleft Logarithmic
potentials with external fields\textquotedblright . Spring, New York-Berlin,
(1997)

\bibitem{Soshnikov} A. Soshnikov. \textquotedblleft Determinantal random
point fields\textquotedblright . Russian Math. Surveys \textbf{55}:5 923-975
(2000)

\bibitem{T} William F. Trench. "Introduction to Real Analysis", Free Edition
1.03, February 2010

\bibitem{VPM} Alexei M. Veneziani, Tiago Pereira and Domingos H. U.
Marchetti. Paper in preparation

\bibitem{VPM1} A. M. Veneziani, T. Pereira and D. H. U. Marchetti.
"Conformal Universality in Normal Matrix Ensembles", Preprint
ArXiv:0909.3418v1 (2009)

\end{thebibliography}
\end{document}